\documentclass[12pt,a4paper]{article}
\usepackage{graphicx}
\usepackage{float}
\usepackage[utf8]{inputenc}
\usepackage{amsmath,amssymb}
\usepackage{authblk}
\usepackage{hyperref}
\usepackage{geometry}
\title{Bohmian Chaos and Entanglement in a Two-Qubit System}

\author[1]{Athanasios C. Tzemos\thanks{Correspondence: atzemos@academyofathens.gr}}
\author[1]{George Contopoulos}
\author[1,2]{Foivos Zanias}

\affil[1]{Research Center for Astronomy and Applied Mathematics of the Academy of Athens, Soranou Efessiou 4, GR-11527 Athens, Greece. Emails: gcontop@academyofathens.gr (G.C.), foivos.zanias@student.uva.nl (F.Z.)}
\affil[2]{Institute of Physics, University of Amsterdam, Science Park 904, 1098 XH Amsterdam, The Netherlands}

\begin{document}

\maketitle

\begin{abstract}
We study in detail the critical points of Bohmian flow, both in the inertial frame of reference (Y-points) and in the frames centered at the moving nodal points of the guiding wavefunction (X-points), and analyze their role in the onset of chaos in a system of two entangled qubits. We find the distances between these critical points and a moving Bohmian particle at varying levels of entanglement, with particular emphasis on the times at which chaos arises. Then, we find why some trajectories are ordered, without any chaos. Finally, we examine numerically how the Lyapunov Characteristic Number ($LCN$) depends on the degree of quantum entanglement. Our results indicate that increasing entanglement reduces the convergence time of the finite-time $LCN$ of the chaotic trajectories toward its final positive value.
\end{abstract}

\noindent \textbf{Keywords:} chaos; Bohmian quantum mechanics; entanglement

\section{Introduction}\label{sec1}
Bohmian Quantum Mechanics (BQM) is an alternative interpretation of Quantum Mechanics, which predicts deterministic trajectories for the quantum particles \cite{debroglie1926a,Bohm,BohmII,holland1995quantum,bohm2006undivided}. These trajectories are guided by the wavefunction $\Psi$ which describes a given quantum system, i.e. a solution of the Schr\"odinger equation, according to a set of first order in time differential equations, the Bohmian equations of motion:
\begin{eqnarray}
m\frac{dx_i}{dt}=\hbar Im\left(\frac{\nabla_i\Psi}{\Psi}\right).
\end{eqnarray}
BQM predicts the same experimental results as the standard Quantum Mechanics (SQM) and has attracted the interest of many authors, both from theoretical and practical standpoint \cite{durr2004bohmian,kocsis2011observing,pladevall2012applied}.

\textbf{Chaos in Quantum Mechanics has remained an open and actively investigated problem for several decades. In the standard (Copenhagen) formulation of quantum theory, the evolution of a system is governed by the unitary propagation of the wavefunction under the Schrödinger equation. However, this framework does not admit a notion of trajectory, making the study of dynamical complexity and chaotic behavior fundamentally different from the classical case. As a result, quantum chaos is typically characterized through indirect, statistical, or information-theoretic indicators  \cite{haake2010quantum, wimberger2014nonlinear,robnik2016fundamental}.  Among the most prominent tools are spectral statistics, such as the distributions of the successive energy levels \cite{bohigas1984characterization}, the spread of a quantum state over a given basis  (participation ratio) and the operator growth, especially in many-body systems, where out-of-time-order correlators (OTOCs) \cite{hashimoto2017out} are used to find how initially localized operators spread under time evolution. While these approaches yield valuable insights, they do not provide a unified or universally accepted definition of chaos, nor do they offer a trajectory-based picture analogous to classical chaotic motion. }

{In contrast, Bohmian Quantum Mechanics provides a deterministic and trajectory-based formulation of quantum theory, where particles follow well-defined paths guided by the quantum wavefunction. This framework introduces a classical-like phase space structure into Quantum Mechanics and allows one to explore quantum dynamics using the full arsenal of tools from nonlinear dynamical systems theory. Bohmian trajectories evolve under a set of first-order, generally non-autonomous and nonlinear differential equations, making them capable of exhibiting both regular and chaotic motion. As such, concepts like Lyapunov exponents,  stable/unstable manifolds etc. become directly applicable in the quantum regime. This makes BQM uniquely suited to studying the dynamical origins of quantum chaos in a physically transparent and geometrically clear manner. Thus, BQM  offers new perspectives and tools for understanding complex quantum behavior in systems ranging from single-particle models to quantum many-body dynamics.}


The origin of Bohmian chaos has been the subject of many works in the past \cite{iacomelli1996regular, frisk1997properties, falsaperla2003motion, wisniacki2005motion,wisniacki2007vortex, borondo2009dynamical,cesa2016chaotic} where it was noticed that chaos is produced due to the interaction of a Bohmian particle with the nodal points of the corresponding wavefunction (where $\Psi=\Psi_R+i\Psi_{I}=0$). However, as we showed in \cite{efth2009}, it is not the nodal point $N$ itself, but its accompanying unstable point in the frame of reference of the nodal point, the `X-point', which is responsible for the generation of chaos. The points $N$ and $X$ form a `nodal point-X-point complex' (NPXPC), a special dynamical structure of the Bohmian flow in the close neighborhood of $N$. Whenever a quantum particle comes close to a NPXPC, its trajectory deviates due to the interaction with the X-point. This interaction is accompanied by a shift of the `finite time Lyapunov characteristic number' ($\chi$). The cumulative effect of many such scattering events is the convergence of $\chi$ to the $LCN$ at a positive value, which is the hallmark of chaos.

In our works we studied different wavefunctions of the unperturbed 2-d quantum harmonic oscillator (which is the most well studied system in the field of Bohmian chaos \cite{makowski2001simplest}) through the prism of the NPXPC mechanism. We note that while this system is classically analytically solvable, i.e. it has no chaos, its Bohmian counterpart has a very rich non-linear dynamics with both ordered and chaotic trajectories. Most results were found in the case of a wavefunction with a single nodal point (see section 2), since in this case we know analytically the position of the nodal point in space for every time. We also studied more complicated wavefunctions with two, three and multiple nodal points scattered around the phase space with different geometries \cite{tzemos2022bohmian}. In this case the position of the nodal points is found, in general, in a numerical way. However, we found a special wavefunction composed of coherent states of the harmonic oscillator, where again we know the position of the nodal points analytically \cite{tzemos2019bohmian,tzemos2020ergodicity}. With a proper choice of physical parameters, this wavefunction resembles a system of two entangled qubits, something very important for applications in quantum information theory \cite{nielsen2004quantum}.

Furthermore, we found that in the above case all chaotic trajectories {are} ergodic. Namely, any two different chaotic trajectories acquire practically the same distribution of points  in the long run, something that we showed in \cite{tzemos2020chaos,tzemos2020ergodicity} by using their `colorplots', where different colors represent the number of trajectory points in a bin of a dense square grid. The distance between the produced colorplots was measured by using the Frobenious distance \cite{strang1993introduction} between the corresponding underlying matrices. Thus chaotic trajectories give the same colorplot. 
On the other hand, the ordered trajectories have the shape of deformed Lissajous figures and give different colorplots, i.e. they are not ergodic.
We note here that in the classical case, ergodicity \cite{coudene2016ergodic} is defined in autonomous systems with a finite phase space. In  BQM the available phase space has, strictly speaking, infinite size but in practice is confined in the `` effective support'' of the wavefunction, i.e. the region where the probability density of the wavefunction is not extremely small (see also  \cite{avanzini2017quantum} for ergodicity in BQM).

In all the above cases we studied in detail the NPXPC mechanism  and its effectiveness in understanding the generation of Bohmian chaos. The main result was that in general chaos emerges from the action of the X-points on the Bohmian trajectories and the NPXPC mechanism accounts for the general profile of the $LCN$. However, there were some minor events in the finite time $LCN$, in which the X-point was far   from the trajectory. These events could not be understood until our work \cite{tzemos2023unstable} where we showed that besides the X-points, there are also unstable fixed points in the inertial frame of reference. We called them `Y-points' and found that they are responsible for these minor contributions to the  $LCN$. The combined action of $X$ and Y-points provides a full explanation of the profile of the $LCN$. On the other hand, if the trajectory of a particle never approaches an X-point or a Y-point then the trajectory is ordered and not chaotic.

Both in the case of a single node wavefunction (there is only one Y-point whose position is found analytically) and in various multinodal wavefunctions we found that, in general, the Y-points are distant from the X-points. But the generality of the above result remained an open problem. As we will show in the present paper, there are cases where the contribution of the Y-points in chaos production is comparable to that of the X-points.

In particular, we will study the case of infinitely many $N,X$ and  Y-points of two entangled qubit states. We will provide the analytical formulae for their positions, in addition to that of the nodal points. As we will show, in this case all the critical points are close to each other at any time $t$. Moreover, every nodal point has two X-points, one ahead of it and one behind it. All these new findings make the study of the onset of  chaos much more complicated. Thus, in order to understand it better,  we will show in detail some representative cases of approaches to the nodal points, X-points and Y-points.

{Entanglement is a fundamental aspect of quantum mechanics, endowing quantum systems with unique properties for information storage, processing, and transmission \cite{horodecki2009quantum}. It is of immense theoretical and technological interest, as it underlies many key developments in quantum computing and communication \cite{erhard2020advances,benatti2020entanglement,chang2023large,chang2025recent}. In the context of Bohmian mechanics \cite{kocsis2011observing,braverman2013proposal,xiao2017experimental,zander2018revisiting,foo2024measurement}, entanglement is essential for inducing complex, potentially chaotic behavior in the trajectories. Without entanglement, the system of Bohmian equations becomes decoupled, and particle dynamics reduce to independent, regular motion. Unlike classical systems, where chaos can emerge from nonlinear interactions between degrees of freedom, entanglement has no classical counterpart. Therefore, studying how entanglement affects chaos in Bohmian trajectories provides a unique window into the quantum origins of dynamical complexity.
We are going to show some results on the relation between entanglement  and the behavior of the $LCN$ for various sets of initial conditions, something that is an open problem in BQM.} 

The structure of the paper is the following: In section 2 we make a short review of the NPXPC mechanism, including the Y-points, in the case of a single nodal point and then pass to the case of two entangled qubits, where we provide the formulae for the positions of the Y-points. Then we study some characteristic trajectories for very small and intermediate values of entanglement and give their distances from $N,X,Y$ at different scattering events as a function of time, discussing their effects on the shape of the trajectories and on the finite time $LCN$. {At the end of section 2 we discuss the ergodic character of the chaotic Bohmian trajectories.}  In section 3 we make a discussion about the ordered trajectories and their different origin depending on commensurability of the frequencies of the oscillator. In section 4 we provide  numerical evidence that quantum entanglement does not affect the value of the $LCN$ itself in a certain way but affects the time of its convergence to a final value. In section 5 we make a summary of our results and draw our conclusions. Finally, in Appendix A we give details about the detection of the X-points and in Appendix B we discuss the periodicity of the trajectories in the case of commensurable frequencies.

\section{The mechanism of chaos in BQM}


As we mentioned in the introduction, the NPXPC is a geometrical structure which characterizes the local geometry of the Bohmian flow near a moving node. The NPXPC changes in time due to the non-autonomous nature of the Bohmian equations of motion. In fact, the distance between $N$ and $X$ decreases as the velocity of the nodal point increases, while  the nature of the nodal point alternates between repeller and attractor  \cite{efth2009}.

\begin{figure}
    \centering
    \includegraphics[width=0.467\linewidth]{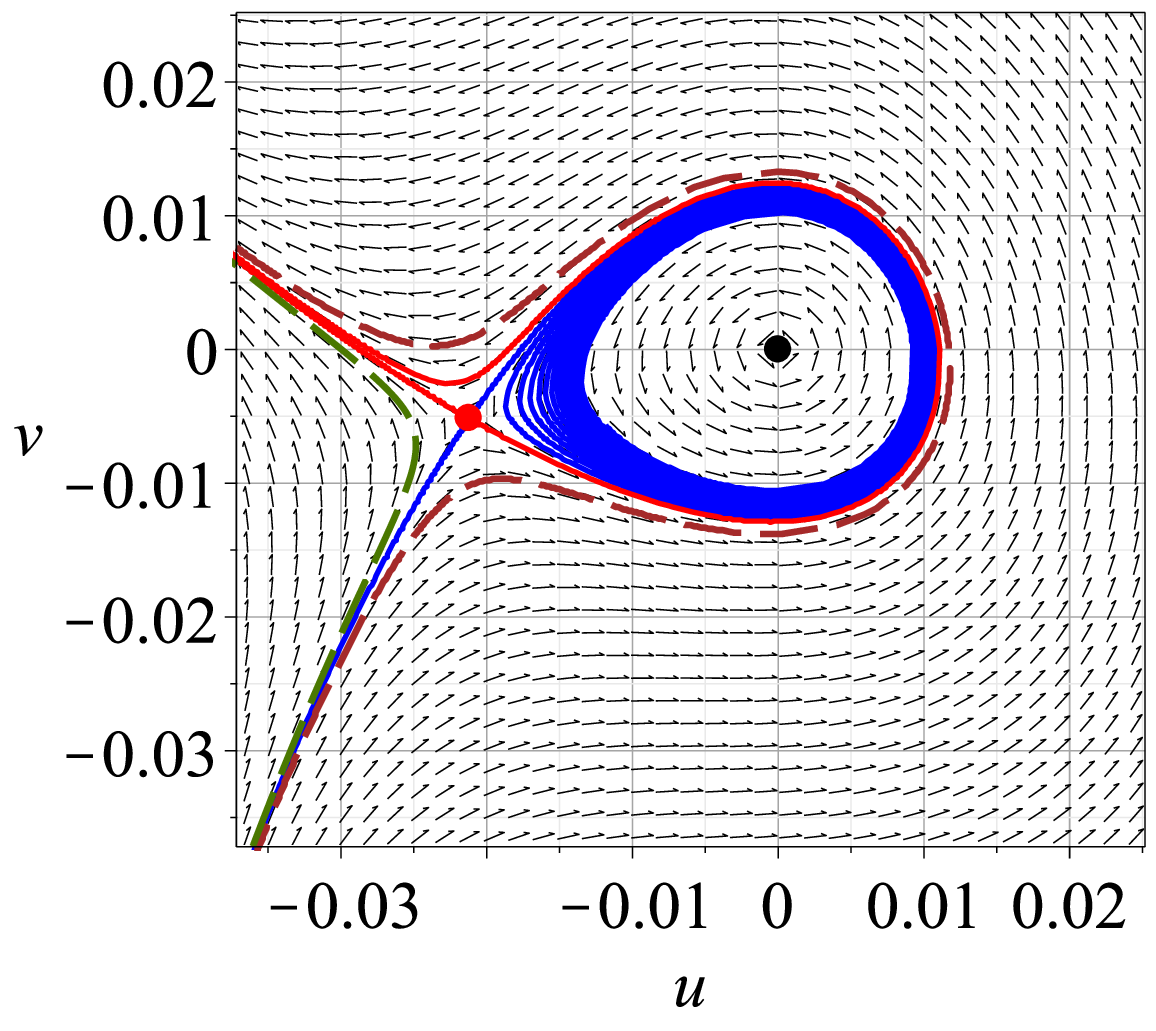}[a]
    \includegraphics[width=0.467\linewidth]{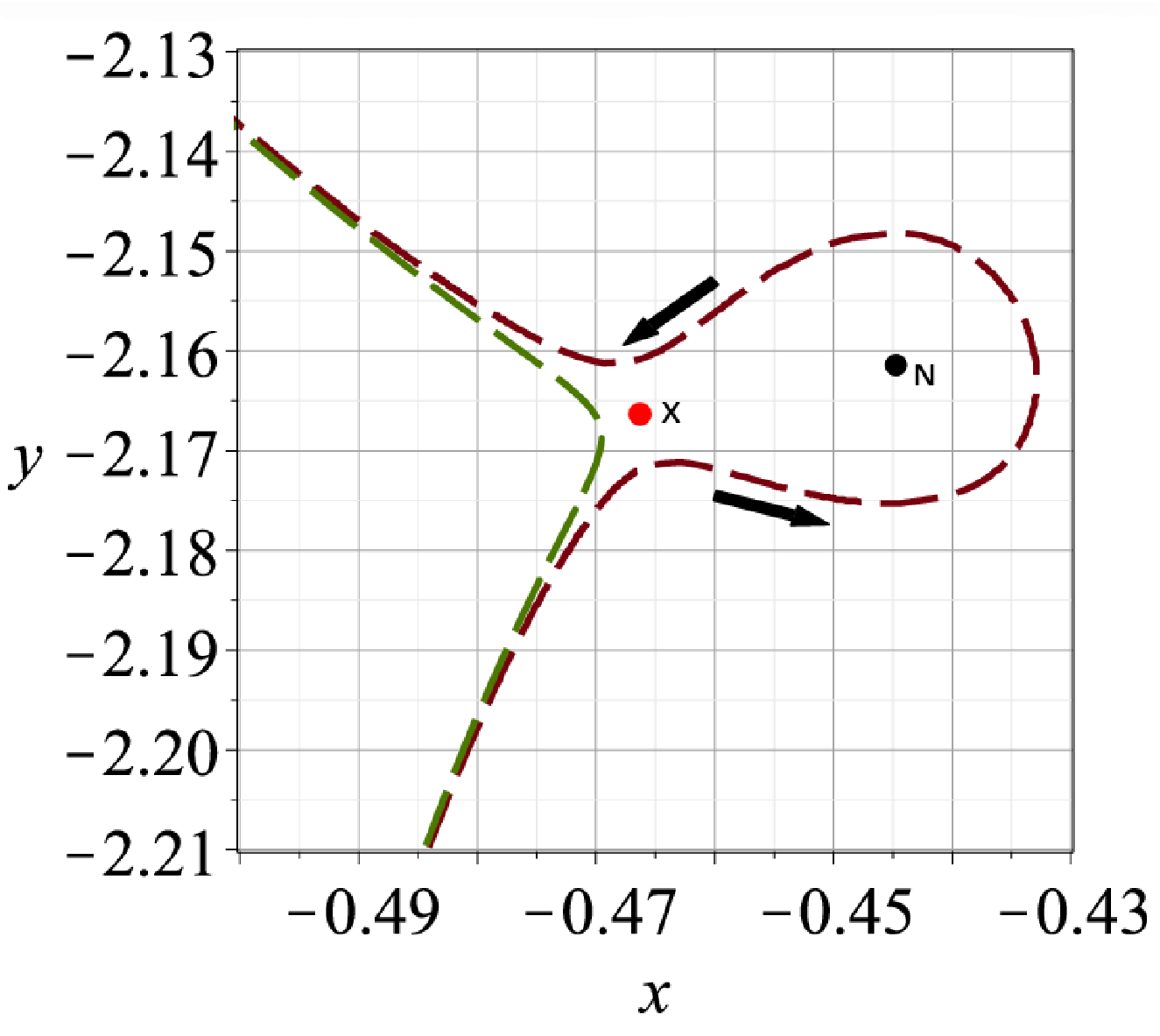}[b]
    \includegraphics[width=0.467\linewidth]{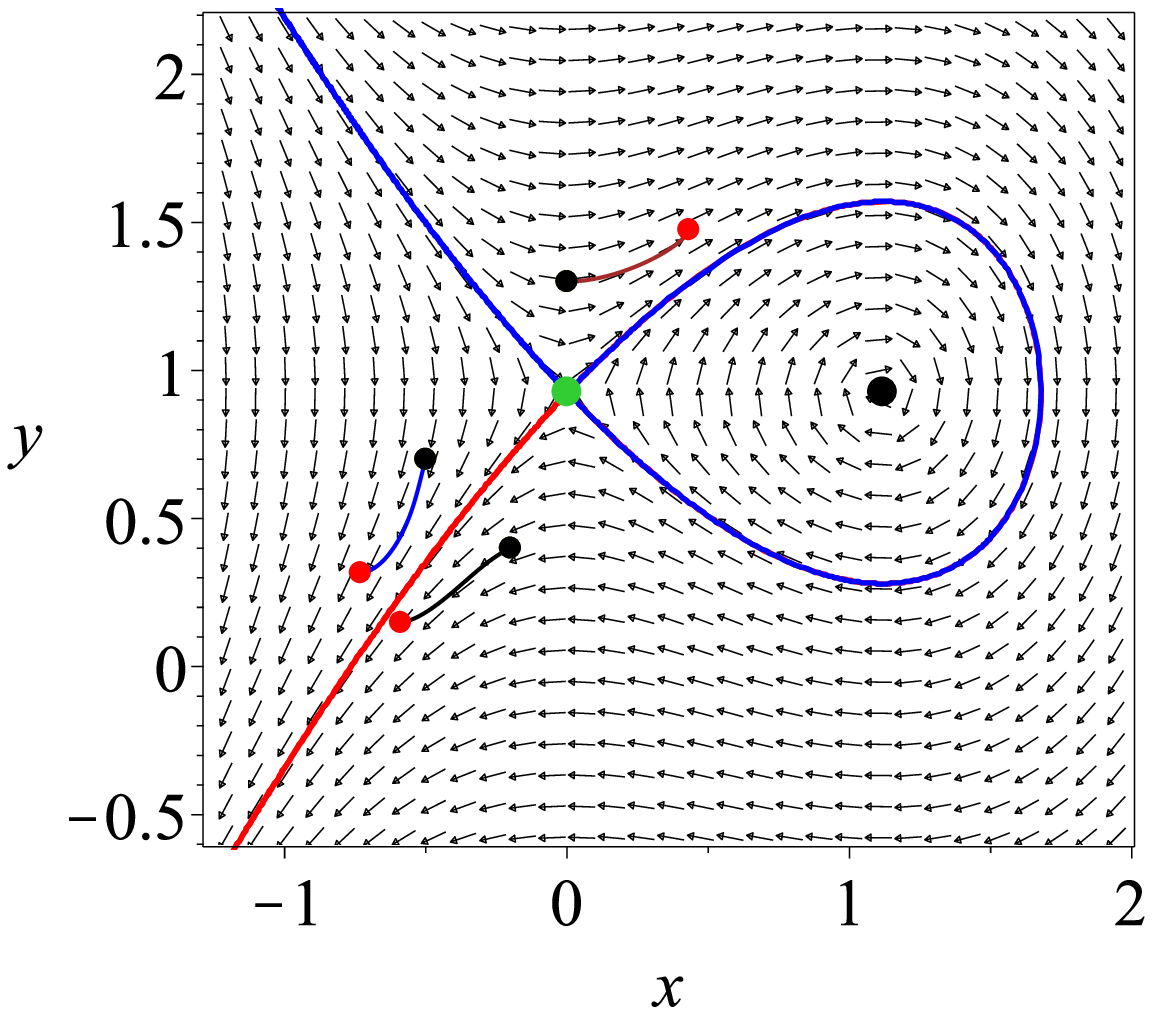}[c]
    \caption{(a) The asymptotic curves of the X-point corresponding to the wavefunction $\Psi=a\Psi_{00}+b\Psi_{10}+c\Psi_{11}$ with $a=b=1, c=\sqrt{2}/2$ at $t=1.5$, in the reference frame of the nodal point ($u = x-x_N$, $v = y - y_N$) (red unstable, and blue stable). The black dot at $u=v=0$ is the nodal point, while the red dot at $u=-0.2123, v=-0.5132$ is the X-point. We plot also two trajectories approaching the X-point and getting deflected by it. b) The same trajectories in the inertial frame of reference and the $X$ and $N$ points at $t=1.5$. c) The corresponding Y-point and its stable (blue) and unstable (red) manifolds in the inertial frame of reference at $t=1.5$ together with arcs of 3 trajectories close to the Y-point. We note that both in a) and c) the blue and the red  asymptotic curves may come arbitrarily close to each other but they never overlap as ensured by the uniqueness theorem of differential equations.}
    \label{fig:triplet}
\end{figure}

{An example of such a NPXPC is presented in Fig.~\ref{fig:triplet}, where we examine the case of a two-dimensional quantum harmonic oscillator corresponding to the classical Hamiltonian}
\begin{equation}\label{ham}
    H = \frac{1}{2}\left( \dot{x}^2 + \omega_x^2 x^2 + \dot{y}^2 + \omega_y^2 y^2 \right)
\end{equation}
{with a wavefunction of the form}
\begin{equation}\label{single}
    \Psi=a\Psi_{0,0}+b\Psi_{1,0}+c\Psi_{1,1}
\end{equation}
where  $\Psi_{m,n}(x,y)=\Psi_m(x)\Psi_n(y)$ and  $\Psi_m(x),\Psi_n(y)$ are the 1-d energy eigenstates the oscillator in $x$ and $y$ coordinates respectively, i.e.
\begin{equation}
\Psi_{m,n}=\prod_{q=x}^y N_q\exp\left(-\frac{\omega_qq^2}{2\hbar}\right)\exp\left(-\frac{i}{\hbar}E_{r}t\right)H_r\left(\sqrt{\frac{M_q\omega_q}{\hbar}}q\right),\label{mn}
\end{equation}
and $r=m,n$ (integers) for $x$ and $y$ respectively  and the normalization constant $N_q=\frac{(M_q\omega_q)^\frac{1}{4}}{\pi\hbar\sqrt{2^rr!}} $. $H_m, H_n$ denote the Hermite polynomials in $\sqrt{\frac{M_x\omega_x}{\hbar}}x$ and   $\sqrt{\frac{M_y\omega_y}{\hbar}}y$ of degrees $m$ and $n$ respectively. Finally, the energy of $\Psi_{m,n}$ is
$E_{m,n}=E_m+E_n=\left(\frac{1}{2}+m\right)\hbar\omega_x+\left(\frac{1}{2}+n\right)\hbar\omega_y.$

{The wavefunction \eqref{single} has been extensively studied in the field of Bohmian chaos since it has only one nodal point whose coordinates are given analytically \cite{parmenter1995deterministic, makowski2001simplest}:}
\begin{equation}
    x_N=-{\frac {a\sqrt {2}\sin \left[ \left(\omega_x + \omega_y\right)t \right] }{2\sqrt {\omega_{x}}\,b\sin \left[ \omega_{y} t\right] }},\quad y_N=-{\frac {b\sqrt{2}\sin \left[ \omega_{x}t \right] }{2 \sqrt {\omega_{y}}\,c\sin \left[ \left(\omega_x + \omega_y\right)t \right]}}
\end{equation}
{simplifying significantly the study of chaos generation with NPXPC mechanism.}

To find the location of the X-point at a fixed time $t$, we begin with the original Bohmian velocity field defined by the system $\frac{dx}{dt}$, $\frac{dy}{dt}$, and pass to a co-moving reference frame centered at the nodal point by defining  new variables $u = x - x_{\text{nod}}$ and $v = y - y_{\text{nod}}$, and transforming the equations to 
\begin{equation}
    \frac{du}{dt} = \frac{dx}{dt} - \frac{dx_{nod}}{dt}, \quad \frac{dv}{dt} = \frac{dy}{dt} -\frac{dy_{nod}}{dt},
\end{equation}
where $\frac{dx_{nod}}{dt}$ and $\frac{dy_{nod}}{dt}$ are the velocities of the nodal point in the  inertial frame of reference. The X-point is then defined as the fixed point of the flow in this moving frame and is determined numerically by solving the equations
\begin{equation}
\frac{du}{dt} = 0, \quad \frac{dv}{dt} = 0.
\end{equation} 

Once the X-point is found, we calculate the Jacobian matrix of the Bohmian flow in the $(u,v)$ coordinates, evaluated at its location. The eigenvalues and eigenvectors of this matrix define the local linear dynamics and provide the stable and unstable eigendirections. To visualize the corresponding asymptotic curves, we integrate the equations of motion by using a fictitious time parameter $s$, associated with an autonomous system obtained by freezing the time dependence of the Bohmian flow at the fixed time $t$ (Fig.~\ref{fig:triplet}a). This approach, known as the `adiabatic approximation', assumes that the time variation of the flow is sufficiently slow in the neighborhood of $t$, allowing us to study the local scattering of Bohmian trajectories by the NPXPC as if the structure were momentarily frozen. The trajectories starting close to the lower stable asymptotic curve approach the X-point and deviate to the left or to the right close to the unstable asymptotic curves of $X$ (Fig.~\ref{fig:triplet}a) in the frame of reference $(u,v)$ centered at the nodal point $N$. In the inertial frame of reference $(x,y)$ these orbits are shown in Fig.~\ref{fig:triplet}b, while $N$ and $X$ have  particular positions at the time $t$.

In this case, we see that the  nodal point acts   as a repeller. The unstable asymptotic curves of the X-point, which govern the  divergence of the trajectories in forward time and are shown in red, while the stable ones, which attract trajectories toward the X-point, are shown in blue. Some trajectories become temporarily trapped near the nodal point and exhibit spiral motion around it as the nodal point moves, a phenomenon known as a \textit{Bohmian vortex}. This vortex behavior persists while the trajectories remain within the region bounded by the nodal point and the X-point. However,  the trajectories exit from this region as the nodal point accelerates and tends to infinity (when the denominator of $x_N$ or $y_N$ becomes zero).

With the NPXPC mechanism we were able to study a variety of cases and explain in general the form of the time evolution of the finite time Lyapunov characteristic number. However, in \cite{tzemos2023unstable} we showed the existence of a fixed point of the Bohmian flow in the inertial frame of reference, which is also unstable. This point, referred to as the `Y-point', is defined as the solution of the system
\begin{equation}
    \frac{dx}{dt}=\frac{dy}{dt}=0.
\end{equation}
{In the case of the wavefunction with a single node \eqref{single} the position of the Y-point is found analytically:}
\begin{equation}
    x_Y = 0, \quad y_Y = -\frac{b \sqrt{2} \sin(\omega_x t)}{2c \sqrt{\omega_y} \sin\left[ \left(\omega_x + \omega_y\right)t \right]}.
\end{equation}
We note that in this case the Y-point has the same values of $y$ as the nodal point, i.e. $y_Y=y_N$. In this frame of reference ($x$, $y$) the asymptotic curves of the Y-point are shown in Fig.~\ref{fig:triplet}c for a fixed time $t$. {We observe that they come so close that they appear to overlap, which would violate the uniqueness theorem for solutions; however, they never actually intersect.}

{In the above case the }node $N$ and its accompanying X-point are in general far from the Y-point. {Thus the latter} was found to play a secondary role in the generation of chaos \cite{tzemos2023unstable}. Similar results were found in other wavefunctions of the 2-d quantum harmonic oscillator \eqref{ham} with two or more nodal points randomly located on the $x-y$ plane.

Another important case that we studied in the quantum analogue of a classical  2-d harmonic oscillator is that of entangled qubits made by coherent states along the $x$ and $y$ axes. 

{We remind that a one-dimensional coherent state in $x$ direction is a special quantum state of the quantum harmonic oscillator, characterized by minimum uncertainty. As such, its evolution closely resembles that of a classical harmonic oscillator. Technically, a coherent state $|\alpha(t)\rangle$ is defined as the eigenstate of the annihilation operator $\hat{\alpha}$, associated with the (generally complex) eigenvalue $\alpha(t)$}:
\begin{equation}
\hat{\alpha}|\alpha(t)\rangle = \alpha(t)|\alpha(t)\rangle,
\end{equation}
where $\alpha(t) = \alpha_0 e^{-i\omega t}$ is a complex eigenvalue, since $\hat{\alpha}$ is non-Hermitian \cite{garrison2008quantum}. 

{When expanded in the basis of the energy eigenstates, the corresponding wavefunction is}
\begin{equation}\label{fock}
Y(x,t) = e^{-|\alpha|^2/2} \sum_{n=0}^{\infty} \frac{\alpha(t)^n}{\sqrt{n!}} \psi_n(x),
\end{equation}
{where $\psi_n(x)$ are the usual energy eigenfunctions of the quantum harmonic oscillator with the corresponding Hermite polynomials }\cite{ballentine2014quantum}
\begin{equation}\label{summing}
\psi_n(x) = \left( \frac{1}{2^n n!} \right)^{1/2} 
\left( \frac{M_x\omega_x}{\pi\hbar} \right)^{1/4} 
e^{- \frac{M_x\omega_x^2}{2\hbar}} 
H_n\left( \sqrt{ \frac{M_x\omega_x}{\hbar} } x \right).
\end{equation}
{This expansion reflects the fact that coherent states are superpositions of all number states, weighted by a Poissonian distribution centered around $|\alpha|^2$.}

{By using the properties of the Hermite polynomials and summing over $n$ in \eqref{summing}, one obtains the wavefunction in the position representation:}
\begin{equation}
Y(x,t) = \Bigg(\frac{m\omega}{\pi\hbar}\Bigg)^{\frac{1}{4}}
\exp\Bigg[-\frac{m\omega}{2\hbar}\Bigg(x - \sqrt{\frac{2\hbar}{m\omega}} \mathrm{Re}[\alpha(t)]\Bigg)^2
+ i\Bigg( \sqrt{\frac{2m\omega}{\hbar}} \mathrm{Im}[\alpha(t)] x + \zeta(t)\Bigg)\Bigg],
\end{equation}
where
\begin{eqnarray}
\mathrm{Re}[\alpha(t)] = a_0 \cos(\sigma - \omega t), \,\,
\mathrm{Im}[\alpha(t)] = a_0 \sin(\sigma - \omega t), \,\,
\zeta(t) = \frac{1}{2}\left[a_0^2 \sin\big(2(\omega t - \sigma)\big) - \omega t\right].
\end{eqnarray}

{We now define the following position wavefunctions:}
\begin{eqnarray}
\nonumber Y_R = Y_R(i,t)\equiv Y(i,t;\omega=\omega_i,m=m_i,\sigma=\sigma_i),\quad i = x,y, \\
Y_L = Y_L(i,t)\equiv Y(i,t;\omega=\omega_i,m=m_i,\sigma=\sigma_i + \pi),\quad i = x,y,
\end{eqnarray}
{which describe one-dimensional coherent states that begin their motion at the \textit{right} or \textit{left} extreme point of the classical oscillation, along the $x$ or $y$ directions respectively. With these, we construct a two-dimensional entangled wavefunction:}
\begin{equation}\label{mult}
\Psi = c_1 Y_R(x,t)Y_L(y,t) + c_2 Y_L(x,t)Y_R(y,t),
\end{equation}
{which displays quantum entanglement between the $x$ and $y$ degrees of freedom (for systems with similar wavefunction see \cite{ghose2001bohmian} . However, since the inner product between two arbitrary coherent states is generally \textit{non-zero}, care must be taken when interpreting this as a qubit system. In the case of a \textit{common amplitude} $a_0$ along both directions, and with $\sigma_x = \sigma_y = 0$, the inner product between $Y_R$ and $Y_L$ becomes $\langle Y_R | Y_L \rangle = \exp(-2a_0^2)$. Thus, for $a_0 = 5/2$, the overlap becomes negligibly small (of order $10^{-6}$), making the two states practically orthogonal. Consequently, the wavefunction \eqref{mult} can be seen as a superposition of two nearly orthogonal product states, thereby effectively realizing a system of {two entangled qubits} (for further details, see \cite{tzemos2019bohmian,zander2018revisiting})}. From now on we work with $m=\hbar=1$.

{The above system besides its technological applications, mainly in quantum optics \cite{garrison2008quantum} is very useful for the study of Bohmian chaos since:}
\begin{enumerate}
    \item {It has infinitely many nodal points. This is due the infinitely many energy eigenstates inside the coherent states \eqref{fock}. The positions of the nodal points are given analytically }
    \begin{eqnarray}
\label{xnod}&x_{nod}={\frac {\sqrt {2}
\left( k\pi\,\cos \left( 
\omega_{y}\,t \right) +\sin \left( 
\omega_{y}\,t \right) \ln  \left( 
\left| {\frac {c_{1}}{c_{2}}} \right|  
\right)  \right) }{4\sqrt {\omega_{x}}a_{0}\,\sin \left(  
\omega_{xy}  t \right) } },\\&
\label{ynod}y_{nod}={\frac {\sqrt {
2} \left(k\pi\, \cos \left( \omega_{x}t 
\right) +\sin \left( \omega_{x}t \right) 
\ln  \left(  \left| 
{\frac {c_{1}}{c_{2}}} \right|  \right)  
\right) }{4\sqrt {\omega_{y}}a_{0}\,\sin 
\left( \omega_{xy}\,t \right) }},
\end{eqnarray}
with $k\in Z $, $k$ even for $c_1\cdot c_2<0$
or odd for $c_1\cdot c_2>0$ and $\omega_{xy}\equiv \omega_x-\omega_y$. 
\item A{s shown in \cite{tzemos2019bohmian} this model has very rich Bohmian Dynamics. In the case of commensurable frequencies all trajectories are periodic, while for non commensurable frequencies we observe the coexistence of order and chaos. From now on we choose to work with positive $c_1,c_2$, $\omega_x=1$, and $\omega_y=\sqrt{3}$ (i.e. with irrational frequencies). Therefore $k = ..., -5, -3, -1, 1, 3, 5, ...$}.
\item {The amount of entanglement in this system can be analytically calculated \cite{tzemos2019bohmian}. Entanglement is maximized when $c_1=c_2=\sqrt{2}/2$ while entanglement is zero for $c_1\cdot c_2=0$}.
\end{enumerate}

{A new result of the present work is that if we calculate the fixed points of the Bohmian flow in the inertial frame of reference we find that there are also infinitely many Y-points between the successive nodal points, whose positions are given analytically by the equations:}
\begin{eqnarray}
\label{xY}&x_{Y}={\frac {\sqrt {2}
\left(2 k'\pi\,\cos \left( 
\omega_{y}\,t \right) +\sin \left( 
\omega_{y}\,t \right) \ln  \left( 
\left| {\frac {c_{1}}{c_{2}}} \right|  
\right)  \right) }{4\sqrt {\omega_{x}}a_{0}\,\sin \left(  
\omega_{xy}  t \right) } },\\&
\label{yY}y_{Y}={\frac {\sqrt {
2} \left(2k'\pi\, \cos \left( \omega_{x}t 
\right) +\sin \left( \omega_{x}t \right) 
\ln  \left(  \left| 
{\frac {c_{1}}{c_{2}}} \right|  \right)  
\right) }{4\sqrt {\omega_{y}}a_{0}\,\sin 
\left( \omega_{xy}\,t \right) }},
\end{eqnarray}
with $2k'=\dots -4,-2,0,2,4\dots$



\begin{figure}[H]
    \centering
    \includegraphics[width=0.4\textwidth]{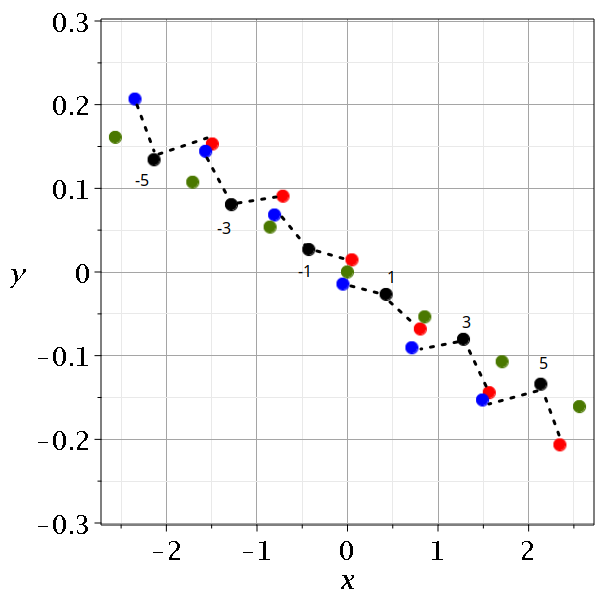}[a]
    \includegraphics[width=0.52\textwidth]{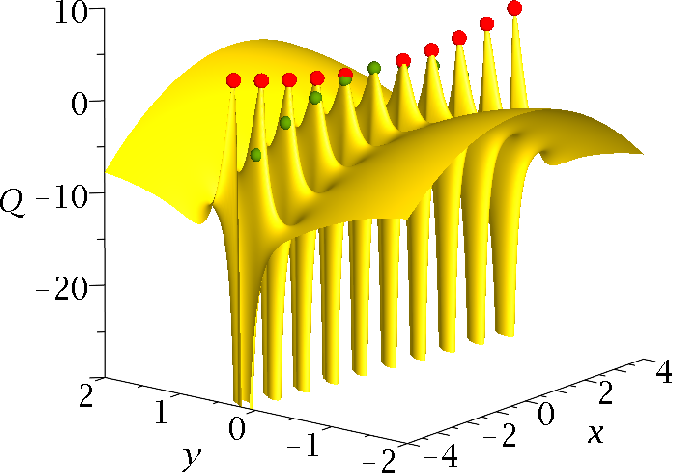}[b]
    \caption{a) The  critical points $N$ (black), $X$ 
 (blue), and $Y$ (green) of the Bohmian flow on the $x-y$ plane at $t=1.5$ for $c_2=\sqrt{2}/2$ and $\omega_x=1, \omega_y=\sqrt{3}$. b) The corresponding 3d surface of the quantum potential $Q$ in the region of a line containing nodal points. The nodal points are at $Q_N=-\infty$. The red points are the X-points while the green points are the Y-points. We observe that every X-point is practically on the top of the local maximum close to a node \cite{tzemos2022bohmianq}. The origin $(0,0)$ corresponds to a Y-point.}
    \label{fig:potential}
\end{figure}

Therefore the Y-points are in the middle of two successive nodal points. This is seen in Fig.~\ref{fig:potential}, where we give the positions of the nodal points, the X-points and the Y-points at a particular time $t=1.5$.  Near every  nodal point $N$ there are two X-points, one on the right (red) and one on the left (blue). These points are close to the Y-points between two successive nodal points (Fig.~\ref{fig:potential}a). In the Appendix A we describe how the X-points are calculated. In Fig.~\ref{fig:potential}a we see that the red X-point (on the right of a nodal point $N$) is very close to the blue X-point of the previous nodal point.

The positions of the $N$, $X$(red) and Y-points (green) on the quantum potential surface $Q$ \cite{tzemos2022bohmianq} \footnote{Q is defined as $Q=-\frac{\hbar^2}{2m}\frac{\nabla^2|\Psi|}{|\Psi|}$, where $\hbar$ is Plancks constant, m is the mass and $\Psi$ is the wavefunction \cite{bohm1984measurement, goldstein2014quantum, dennis2015bohm, hojman2021bohm}. We work with $m=\hbar=1$.} are shown in Fig.~\ref{fig:potential}b. At the origin ($x=0,y=0$) we have a Y-point with $k'=0$. As $|k|$ and $|k'|$ increase the Q values of the red points increase and those of the Y-points decrease. Further away from the lines of the nodal points, the X-points and the Y-points the values of $Q$ increase smoothly. {We observe that beside the central Y-point lying at the origin of the $x-y$ plane the X-points are located at higher values of $Q$ than the Y-points. This shows that in this model we have two kind of unstable points that produce chaos. However, even in this case the total force  $F=-\nabla(V+Q)$ acting on a Bohmian particle is larger at the X-poins than at the Y-points.}

The case of entangled  qubits has a new interesting feature regarding Bohmian chaos: its Y-points are always close to the X-points, something that was not the case in the wavefunction of a single nodal point. This means that both the X-points and the Y-points contribute practically equally in producing chaos.

\begin{figure}[H]
    \centering
    \includegraphics[width=0.4\linewidth]{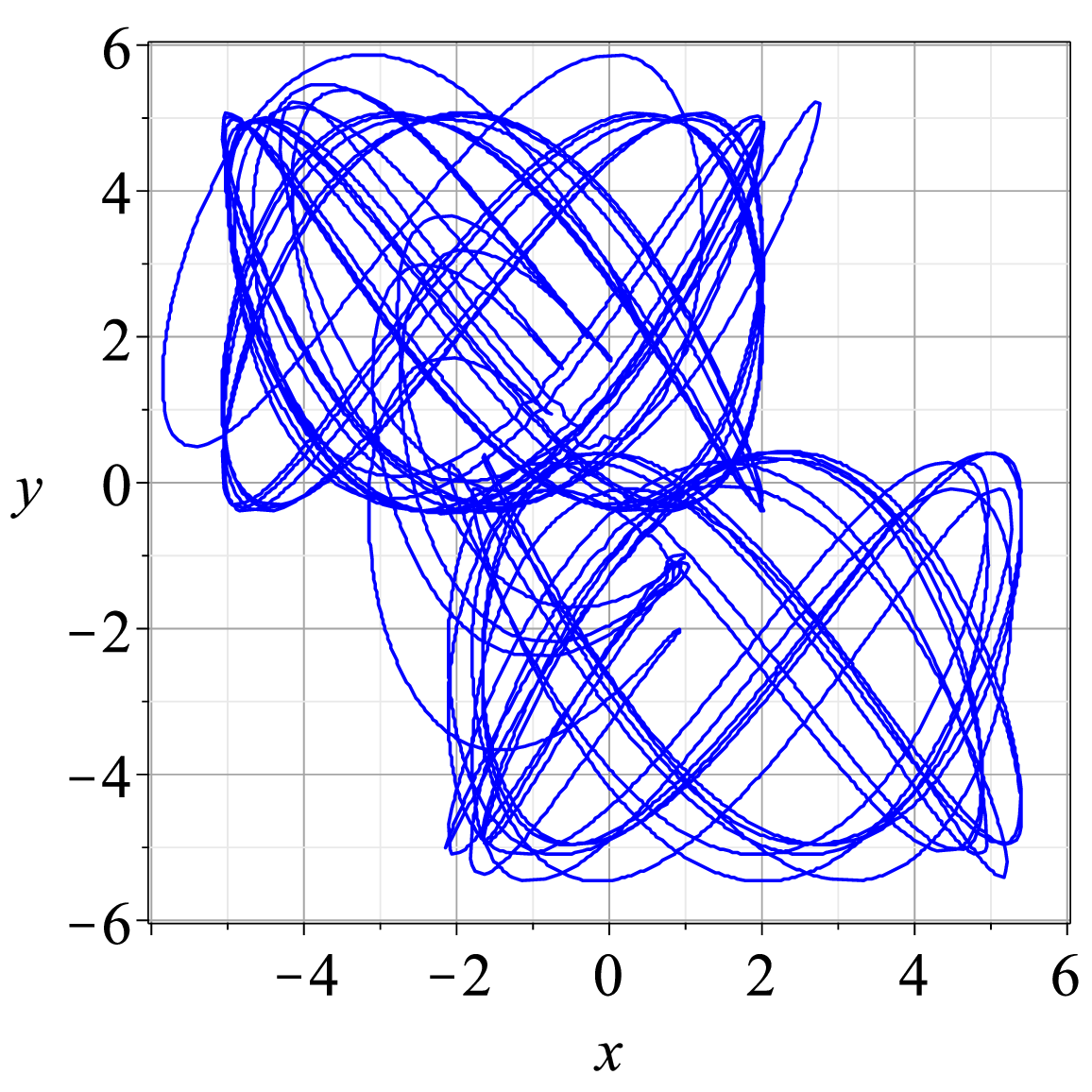}[a]
    \includegraphics[width=0.4\linewidth]{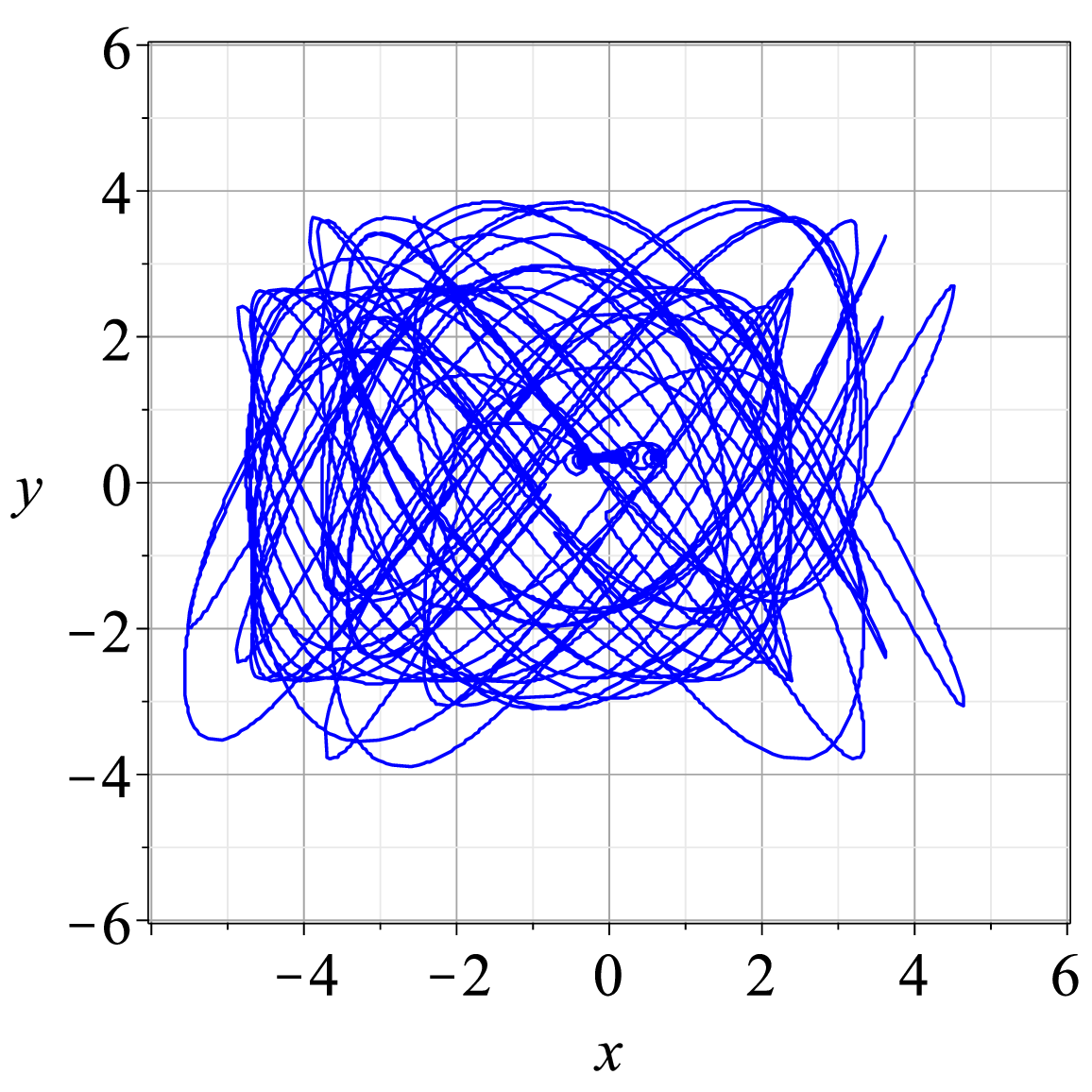}[b]
    \caption{Trajectories of particles in the inertial frame $x, y$ for a time interval $t \in [0, 200]$. (a) For $c_2=0.001$ with $x(0)=0, y(0)=3$. (b) For $c_2=0.3$ with $x(0)=-2.5654, y(0)=3.6585$ . }
    \label{fig:troxies}
\end{figure}

In Fig.~\ref{fig:troxies} we show two characteristic trajectories for $t\in[0,200]$ for different degrees of the entanglement parameter a) with $c_2=0.001$ (weak entanglement) and b) $c_2=0.3$ (moderate entanglement).

If there is no entanglement ($c_2 = 0$), the trajectories in the $x-y$ plane are Lissajous figures (all critical points are at $\infty$). If the entanglement is small, the trajectories form  Lissajous-like figures for some intervals of time. When the particle is far from all nodal points and from the $X$ and $Y$ points, the trajectory is very close to a Lissajous figure, but when it approaches the $N$, $X$ and Y-points, the trajectory undergoes a change. Some close approaches lead to very different Lissajous-like forms (Figs.~\ref{fig:troxies}a,b). But when $c_2$ is close to its maximum ($c_2=\sqrt{2}/2$), the trajectories have no time to form any Lissajous-like forms.

\subsection{A typical Bohmian chaotic trajectory}

\begin{figure}[H]
    \centering
    \includegraphics[width=0.4\linewidth]{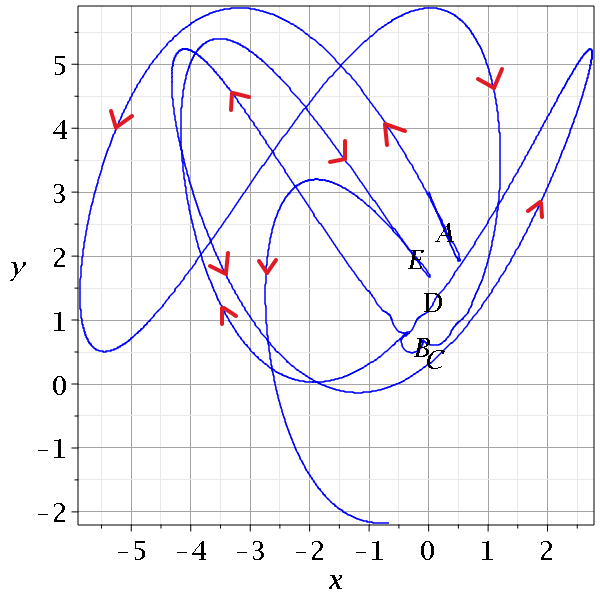}[a]
    \includegraphics[width=0.8\linewidth]{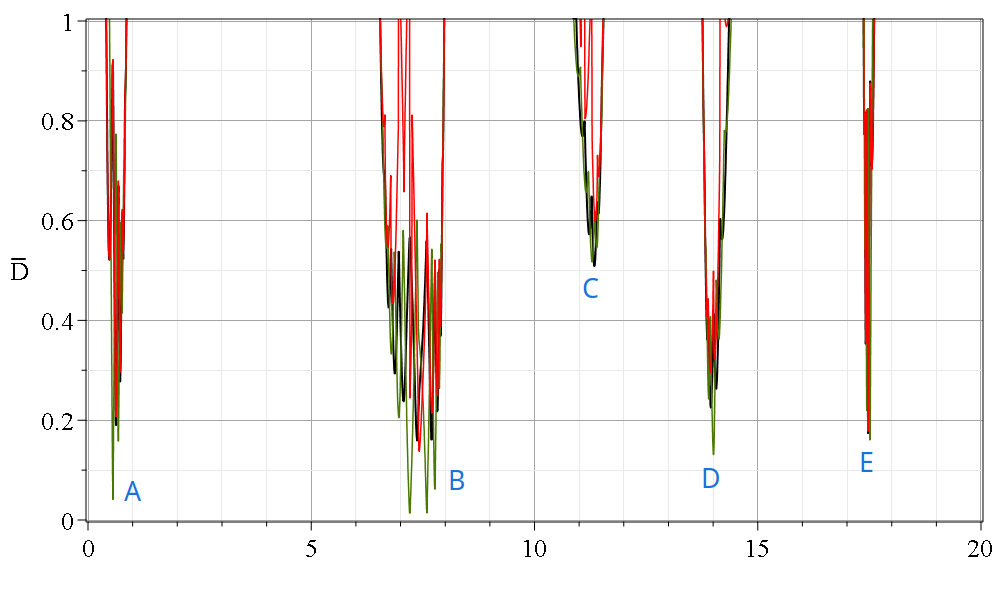}[b]\\
    \includegraphics[width=0.8\linewidth]{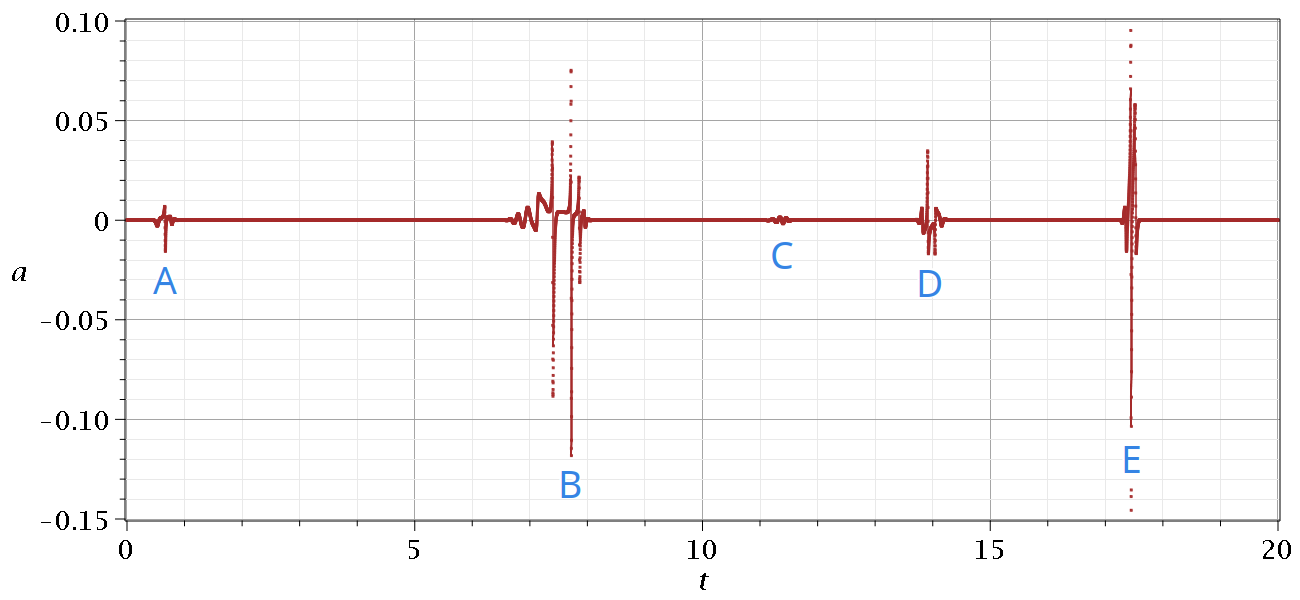}[c]
    \caption{a) A Bohmian trajectory of a particle starting with $c_2=0.001, \omega_x=1, \omega_y=\sqrt{3}$ at $x(0)=0, y(0)=3$ up to $t=20$. b) The distance between the particle and the closest nodal point (black), the closest X-point (red) and the closest Y-point (green) for the same time interval. c) The variations of the stretching number $a$. We observe that its spikes take place when the distances between the critical points and the particle  are minimized. These happens at the points $A-E$ shown on the trajectory.}
    \label{fig:troxia0001}
\end{figure}

In Fig.~\ref{fig:troxia0001} we give the details of some approaches of a Bohmian particle to $N,X,Y$ in the case $c_2 = 0.001$ for a time interval $t \in [0, 20]$. The trajectory is given in Fig.~\ref{fig:troxia0001}a, and in Fig.~\ref{fig:troxia0001}b we give the distances from the closest nodal point (black curve), the closest X-point (red curve) and the closest Y-point (green curve) at every time. The approaches take place near the points $A$, $B$, $C$, $D$ and $E$ of the trajectory (Fig.~\ref{fig:troxia0001}a). We note that near the time of an approach to a particular nodal point $N$, we have also approaches to some nearby $N$, $X$ and $Y$ points. We call such a set of approaches an ``event''. The main events are $A$, $B$, $D$ and $E$ at minimum distances of the order $\tilde{D} = 0.2$, while during the event $C$ the minimum distance is close to $\tilde{D} = 0.6$.

In Fig.~\ref{fig:troxia0001}c we give the corresponding variations of the stretching number, which lead to chaos.
We remind that the stretching number is related to the Lyapunov characteristic number: in particular, if we take two  nearby trajectories at $t=t_0$ and their deviation vector $\xi_k$ at the times $t=s\Delta t, s=1,2,\dots$ then we define the `finite time Lyapunov characteristic number', $\chi$ \cite{efth2009} as 
\begin{equation}\label{chi}
\chi=\frac{1}{t}\sum_{i=1}^sa_i,
\end{equation}
where 
\begin{equation}
a_s=\ln\left|\frac{\xi_{s+1}}{\xi_s}\right|
\end{equation}
is the `stretching number'. Thus $a/\Delta t$ is a ``one step Lyapunov characteristic number''.  The $LCN$ itself is the limit of $\chi$ at $s\to\infty$ and is zero for ordered trajectories, while it is positive for chaotic trajectories (we work with $\Delta t=0.01)$.

\begin{figure}[H]
    \centering
    \includegraphics[width=0.8\linewidth]{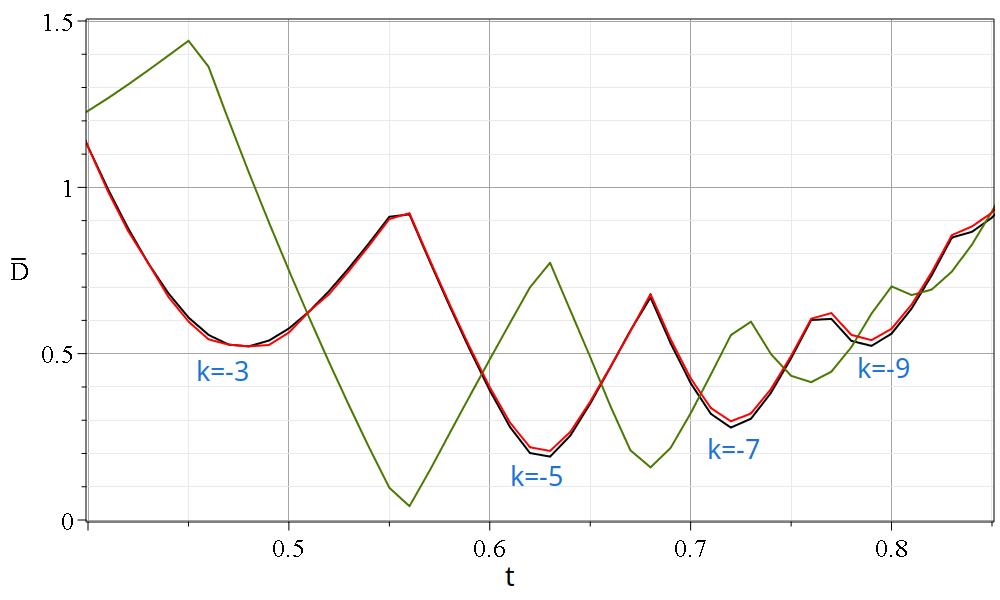}
    \includegraphics[width=0.8\linewidth]{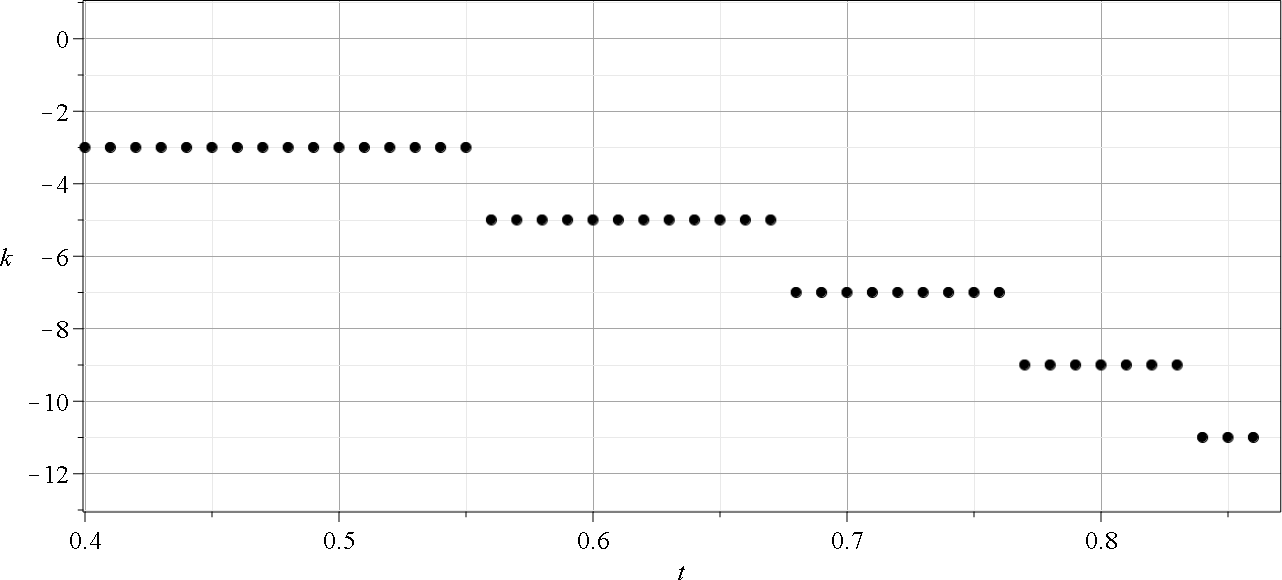}\\
    \includegraphics[width=0.82\linewidth]{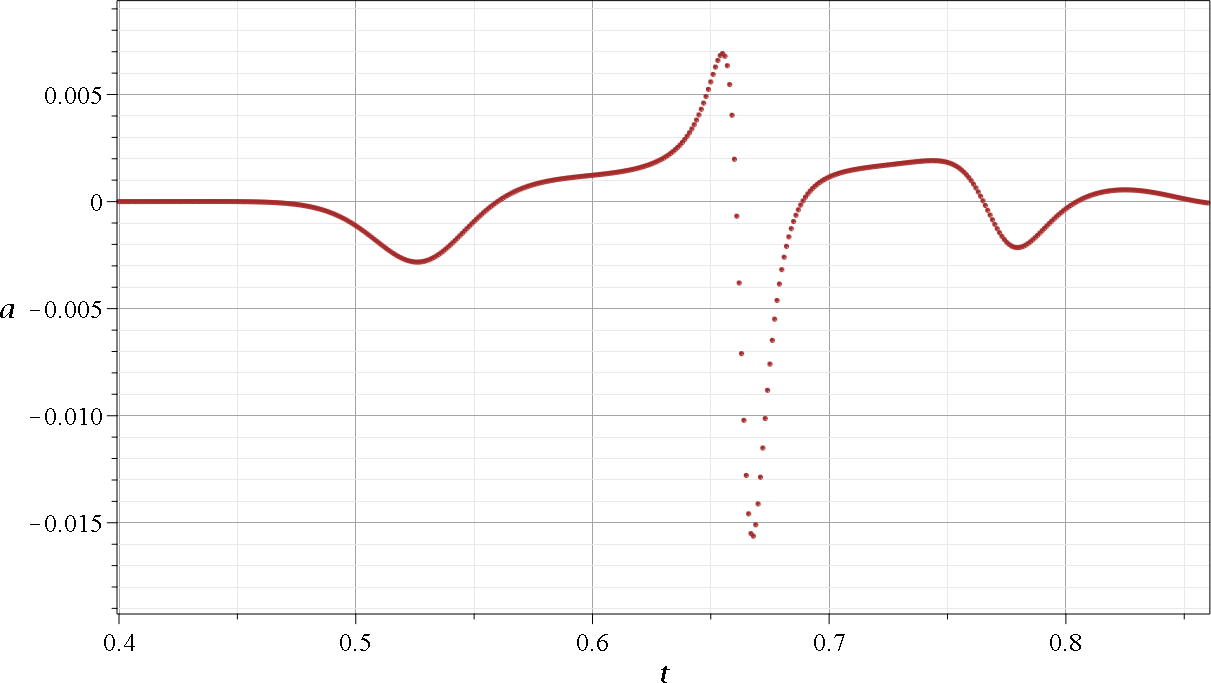}
    \caption{a) Zoom of Fig.~\ref{fig:troxia0001}a at the scattering event $A$ taking place at $t\in[0.4,0.8]$. b) The indices $k$ of the closest nodal point and c) the values of the stretching number $a$  during this time interval.}
    \label{fig:troxiazoom0001}
\end{figure}

We note that the closest points $N$, $X$ and $Y$ to the trajectory of the particle change over time during a single scattering event. This is shown in Figs.~\ref{fig:troxiazoom0001}a,b where we see a zoom in the first event ($A(t\in[0.4-0.85])$). During $A$  the involved nodal points are $k=-3, -5, -7$ and $k=-9$.

The details of Fig.~\ref{fig:troxiazoom0001}a are the following: The particle approaches first (at $t=0.48$) the nodal point $k=-3$ (Eqs.~\ref{xnod}-\ref{ynod}) and beyond that time the distance from this nodal point increases. Then at $t=0.56$ the moving particle comes close to the Y-point (minimum of green curve) with $2k'=-4$, between the nodal points $k=-3$ and $k=-5$. After that time the particle comes closer to the nodal point $k=-5$. Near the time of the minimum distance from the Y-point we have the crossing of the two black curves $k=-3$ and $k=-5$ and of the corresponding X-curves (red). Then at $t=0.62$ the distance from the point $k=-5$ is minimum. At  $t=0.67$ we have a minimum of the green curve ($2k' = -6$)  and a corresponding crossing of the distances from the nodal points $\bar{D}$ (black curves) $k=-5$ and $k=-7$, and the corresponding X-points. Then we have a minimum of the curve $k=-7$, another minimum of the green curve and an approach to the nodal point $k=-9$. Beyond that time all the distances from $N,X,Y$ become large as prior to the event $A$ ($t<0.4$) The closest nodal point for various times is given in Fig.~\ref{fig:troxiazoom0001}b. Therefore during $A$, we have approaches to four nodal points and to three Y-points between them. Similar effects appear if we zoom in the other events $B,C,D,E$.


During an event, we have a number of spikes consisting of increases and decreases of the stretching number $a$ (Fig.~\ref{fig:troxia0001}c). The variations of $a$ during the event $A$ are given in a zoom (Fig.~\ref{fig:troxiazoom0001}c). The value of $a$ is negative and positive during the event, while it is practically zero before and after the event. The total contribution of the values of $a$ to the $LCN$ during this event is slightly negative. However, other events make a positive contribution.

\begin{figure}[H]
    \centering
    \includegraphics[width=0.75\linewidth]{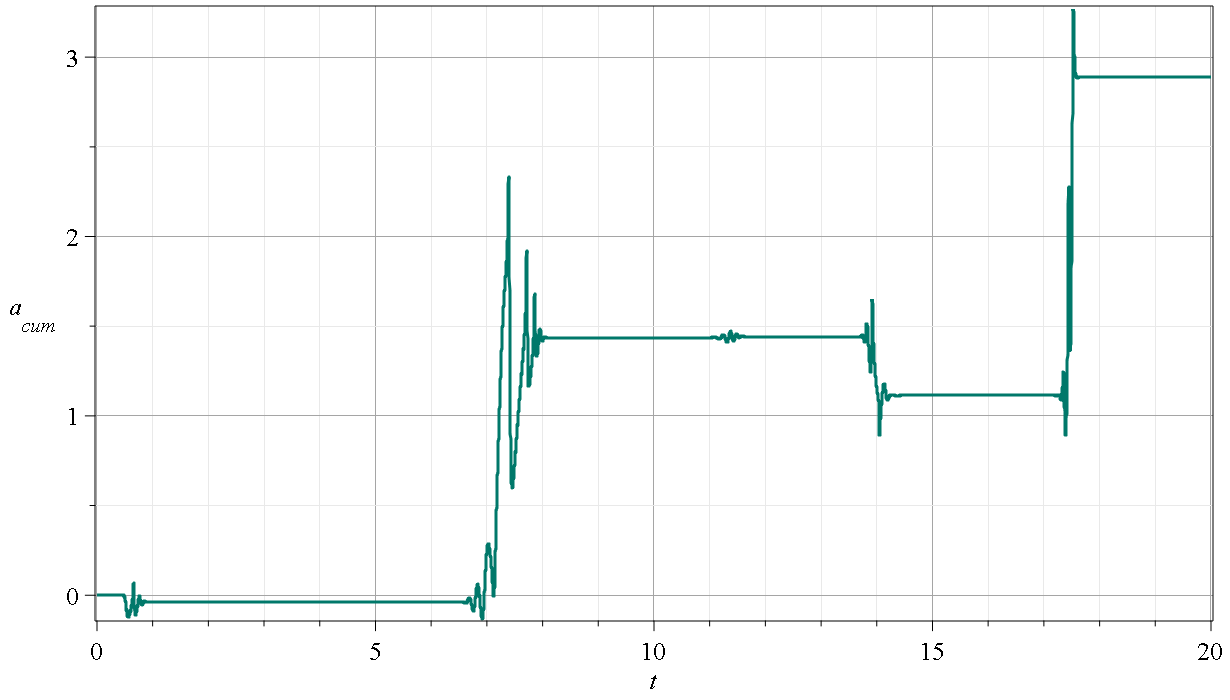}
    \caption{The cumulative stretching number $a_{cum}=\sum a $ in the case $c_2=0.001$ and for $x(0)=0, y(0)=3$ up to 20 time units.  }
    \label{fig:0001acum}
\end{figure}

In Fig.~\ref{fig:0001acum} we give the sum of $a's$ (i.e. the $a_{cum}=\sum a$) (Eq.~\ref{chi}) for times up to $t=20$ \footnote{$a_{cum}$ shows more clearly the net effect of each scattering event on the production of chaos than the stretching number itself.} We see that at every event we have a decrease or an increase of this sum at the successive events, while between
any successive events the sum does not change. In fact, we have decreases at the events $A, D$  increases at the events $B, E$ and practically no change at the effect $C$. The value of the finite time $LCN$, $\chi$, at any time $t$
is the sum divided by the corresponding time (Eq.~\ref{chi}).

Therefore the $acum$ increases on the average. The value of $\chi$ has some fluctuations but it stabilizes at a constant value which is the $LCN>0$ for chaotic trajectories. Therefore chaos is a property that is established after several events and not by a single event (the case of ordered trajectories where $LCN=0$ is discussed in the next section).


During the event $A$ the nodal points are moving and the positions of the nodal points are given at every $\Delta t=0.01$. At $t=0$, all the nodal points are at infinity. As $t$ increases from $t=0$ they come to the central region (around $x=y=0$) for an interval of time (including the duration of the event $A$ from $t=0.4$ to about $t=0.82$) and later at $t=t_\infty=\frac{\pi}{\omega_2-\omega_1}=\frac{\pi}{\sqrt{3}-1}\approx 4.3$ they escape again to infinity. Escapes happen at every multiple of $t_\infty$ and between two successive escapes we have, in general, an ``event" (a set of close approaches to the $N$, $X$ and $Y$ points within a small interval of time).

\begin{figure}[H]
    \centering
    \includegraphics[width=0.45\linewidth]{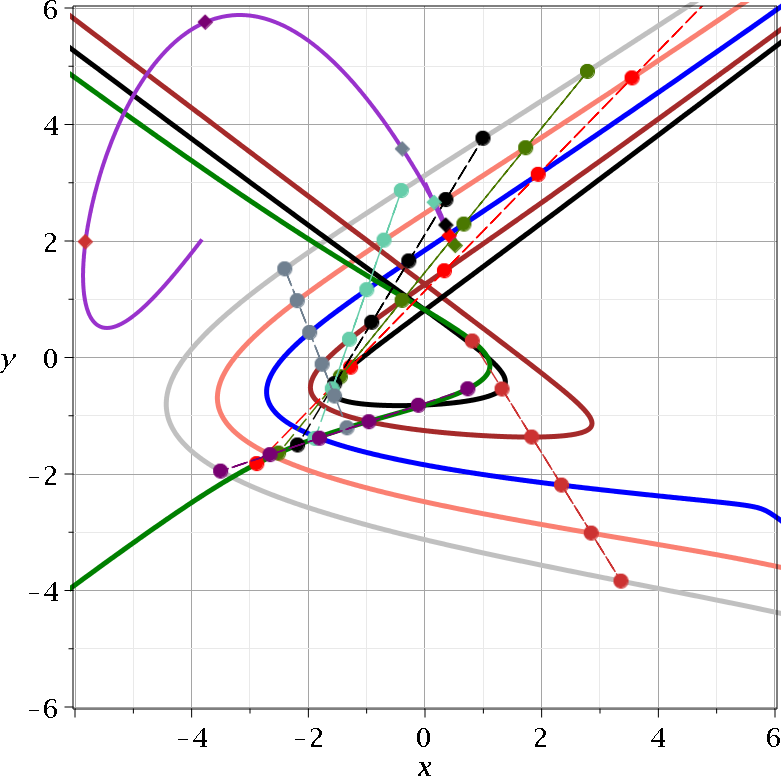}[a]
     \includegraphics[width=0.45\linewidth]{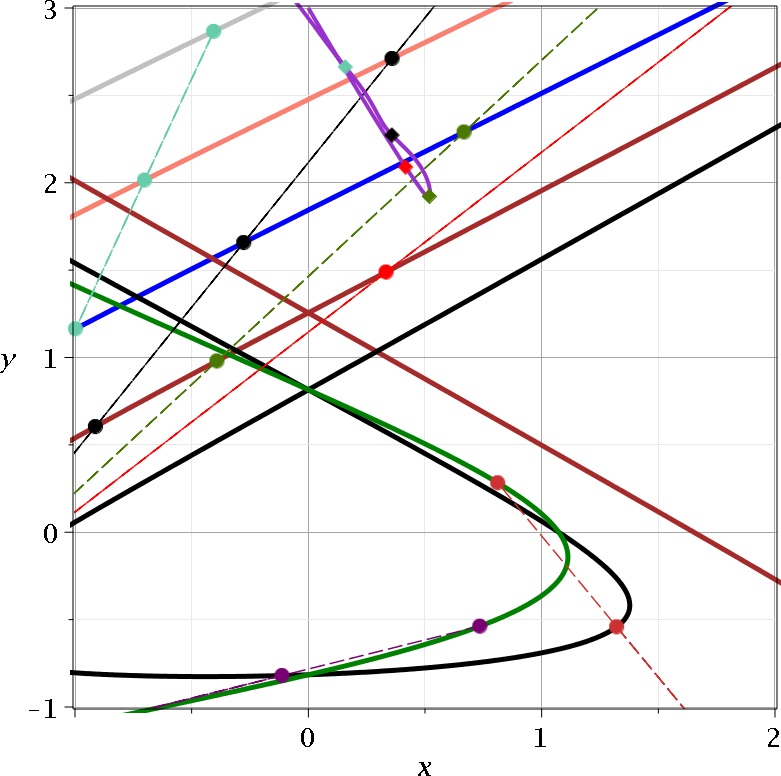}[b]
    \caption{a) The trajectories of the nodal points in the case $c_2=0.001, \omega_x=1, \omega_y=\sqrt{3}$ with $k=1$ (dark green), $k=-1$ (black), $k=-3$ (burgundy), $k=-5$ (blue), $k=-7$ (pink) and $k=-9$ (gray). The nodal points are given at the times $t=0.4$ (red, $t=0.5$ (green), $t=0.6$ (black), $t=0.8$ (cyan), $t=1$ (gray), $t=2$ (dark purple) and $t=3$ (dark orange). We also show a trajectory starting at $x(0)=0, y(0)=3$ for $t\in[0,4.3]$. b) A zoom showing the region where this trajectory comes close to a nodal point at $t\simeq 0.6$ and forms a loop. }
    \label{fig:komboi}
\end{figure}

In Fig.~\ref{fig:komboi}a we give the trajectories of the nodal points for $k=1,-1,-3,-5,-7$ and $-9$. At every time $t$ the nodes are along a straight line. We have marked the lines at $t=0.5$ (red), $t=0.6$ (dark green), $t=0.7$ (black), $t=0.8$ (green), $t=1$ (gray), $t=2$ (yellow) and $t=3$ (red). Initially, the line of nodes makes an angle of about $37^o$ with the $x$-axis. In fact, as $t$ tends to zero, the ratio $y_N/x_N$ tends to $\sqrt{\omega_x/\omega_y}=0.76=\tan{37^o}$, while $x$ and $y$ tend to $+\infty$ for $k<0$, and to $-\infty$ for $k>0$. If $t$ tends to $t_\infty$, the ratio $y_N/x_N$ for large $|k|$ tends to $\sqrt{\frac{\omega_x}{\omega_y}}\frac{\cos{\left(\omega_x t_\infty\right)}}{\cos{\left(\omega_y t_\infty\right)}} = -0.076 = \tan{\left(180^o-37^o\right)} = \tan{143^o}$. The angle of the nodal line with the x-axis increases as $t$ increases from $t=0$, and beyond $t=1$ it becomes larger than $180^o$, and approaches $180^o + 143^o$ as $t\rightarrow t_\infty$. It is of interest to note that the trajectories of the nodal points with $k>0$ start at $x=y=-\infty$ and those with $k<0$ start at $x=y=+\infty$.

The trajectory of a particle starting at ($x=0$, $y=3$) forms a loop during event A, and then the particle moves up and to the left (we stop the calculations in this figure when $t=t_\infty$). The details are shown in Fig.~\ref{fig:komboi}b. The moving particle is at the red point of its trajectory at $t=0.5$, and at the lowest point of the loop for $t=0.6$, when it is next to the nodal point $k=-5$ (compare with Figs.~\ref{fig:troxiazoom0001}a,b). Later it moves upwards.


\subsection{A Bohmian chaotic trajectory with  a vortex}

{Previously we saw chaos generation in a typical Bohmian trajectory. Now we study a trajectory containing a Bohmian vortex, a special phenomenon in Bohmian chaotic dynamics occurring whenever a trajectory comes very close to a moving nodal point and follows its motion forming a spiral around it for a given time interval.}

\begin{figure}[H]
    \centering
    \includegraphics[width=0.5\linewidth]{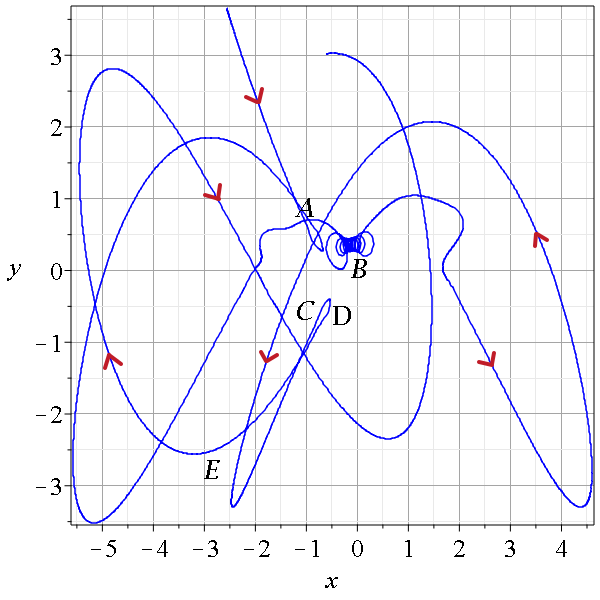}[a]
    \includegraphics[width=0.6\linewidth]{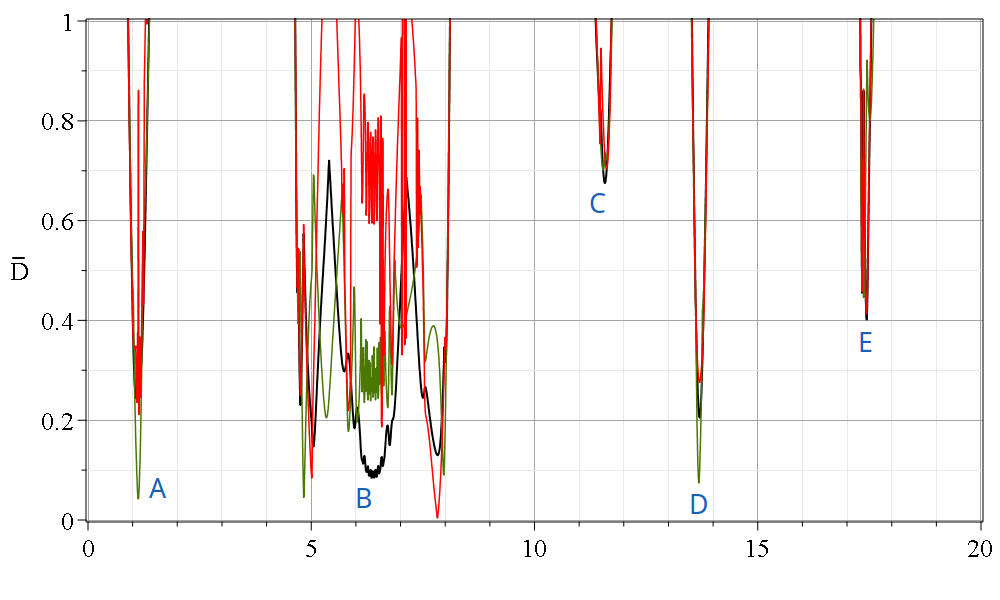}[b]\\
    \includegraphics[width=0.6\linewidth]{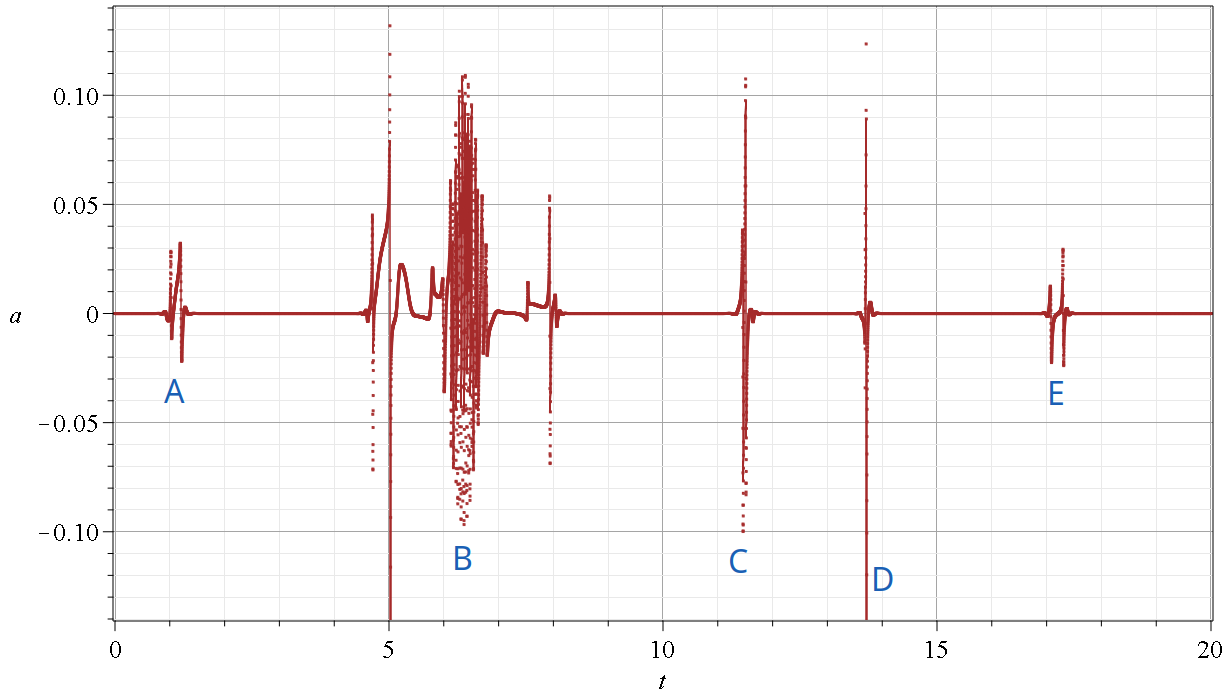}[c]
    \caption{a) A Bohmian trajectory starting at $x(0)=-2.5654, y(0)=3.6585$ up to $t=20$ for $c_2=0.3, \omega_x=1, \omega_y=\sqrt{3}$. b) The distance between the trajectory and the closest nodal point (black), the closest X-point (red) and the closest Y-point (green) for the same time interval. c) The variations of the stretching number $a$.}
    \label{fig:troxia03}
\end{figure}

{In particular, in (Fig.~\ref{fig:troxia03}) we show a  set of events is seen in the case $c_2=0.3$  for $t\in[0,20]$. We have again five events ($A-E$) as in our previous example. But here the event $B$, which corresponds to the Bohmian vortex}, lasts a longer time (Fig.~\ref{fig:troxia03}b,c) than in the previous case with $c_2=0.001$.

\begin{figure}[H]
    \centering
    \includegraphics[width=0.89\linewidth]{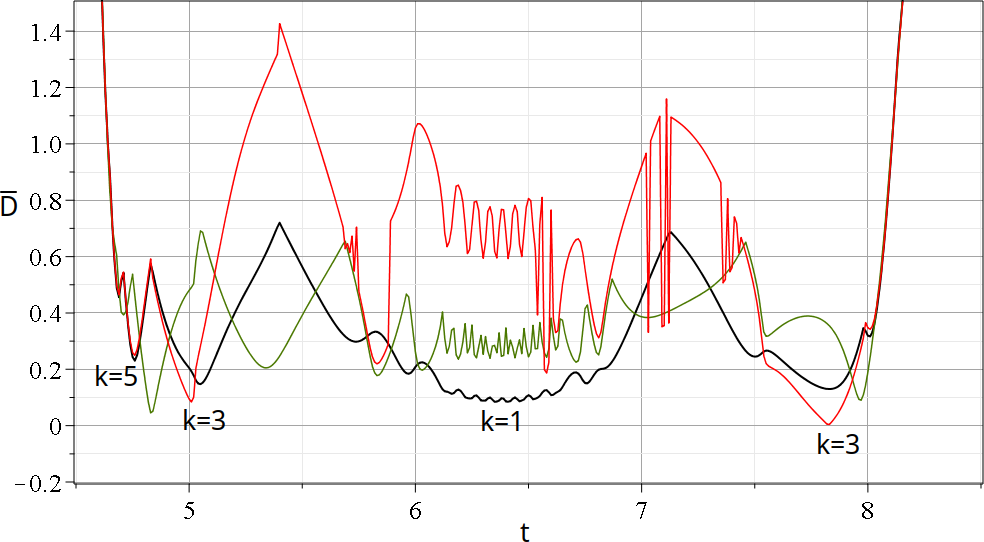}[a]
    \includegraphics[width=0.9\linewidth]{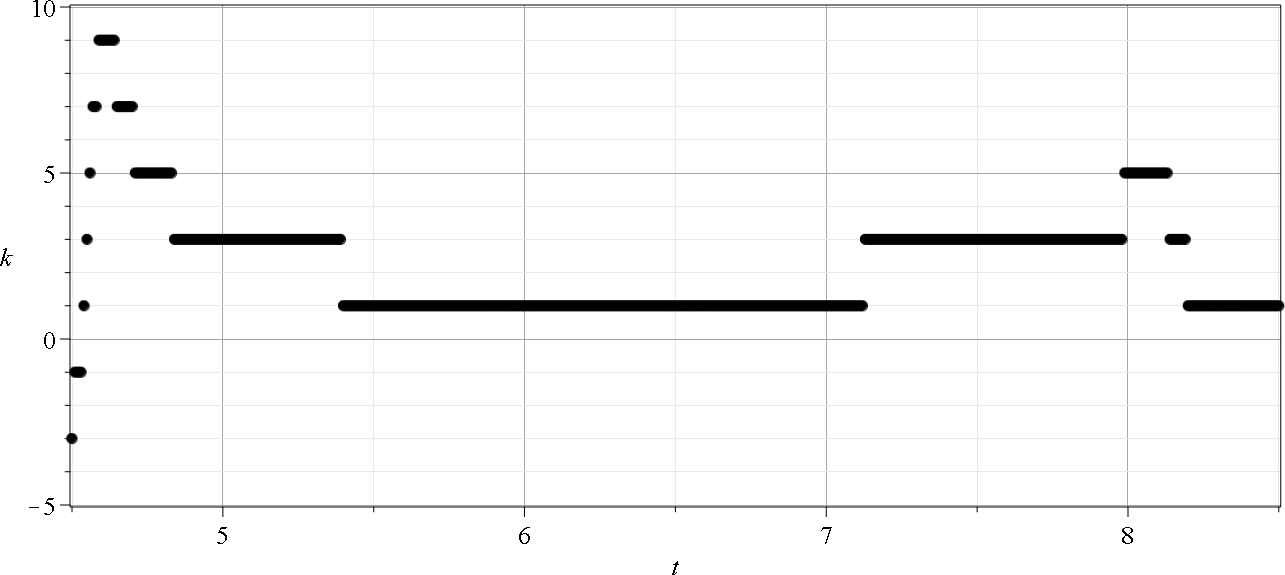}[b]\\
    \includegraphics[width=0.92\linewidth]{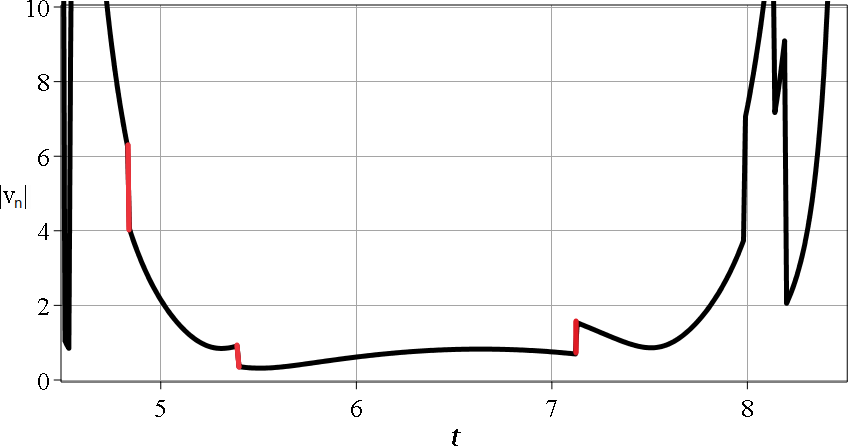}[c]\\
    \caption{a) Zoom of the event $B$ of Fig.~\ref{fig:troxia03}a that includes the Bohmian vortex taking place at $t\in[6,7]$. b) the indices $k$ of the closest nodal point and c) the velocities of the closest nodal point during this time interval.}
    \label{fig:troxiazoom03}
\end{figure}

{More details are given in Fig.~\ref{fig:troxiazoom03}a, where we see that the vortex exists for $t\in[6,6.8]$}. During that time the particle is closer to the nodal point ($k=1$) than to the corresponding X and Y-points. Before the spiral part of the trajectory we had approaches to the nodal points $k=5$ and $k=3$ and to the corresponding $X$ and $Y$ points (Fig.~\ref{fig:troxiazoom03}a,b). After the vortex we have an  approach to the nodal point $k=3$. There are further minima of distances to  other nodal points before $t=4.8$ and after $t=8.1$, but these are at large distances $\tilde{D}$.

The velocities of the nearest nodal point $N$ during the event $B$ are given in Fig.~\ref{fig:troxiazoom03}c. In general, the velocities are relatively large. However, during the approach to the nodal point $k=1$ (where we have the Bohmian vortex) they are quite small. {Moreover $\bar{D}$ shows an oscillatory behavior around the nodal point  (black curve in Fig.~\ref{fig:troxia03}b), in contrast to  the previous example, and because of that, the trajectory undergoes many spikes of the stretching number $a$ during this period (Fig.~\ref{fig:troxia03}c).} In Fig.~\ref{fig:troxiazoom03}c we see three abrupt changes of the velocity at the transitions from the node $k=5$ to $k=3$ (at $t=4.9$) from $k=3$ to $k=1$ (at $t=5.4$) and from $k=1$ to $k=3$ (at $t=7.1$) marked with red segments.

\begin{figure}[H]
    \centering
    \includegraphics[width=0.46\linewidth]{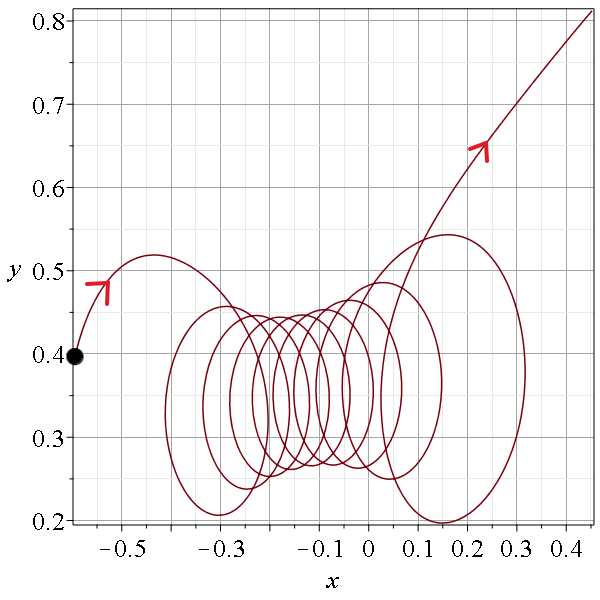}[a]
    \includegraphics[width=0.46\linewidth]{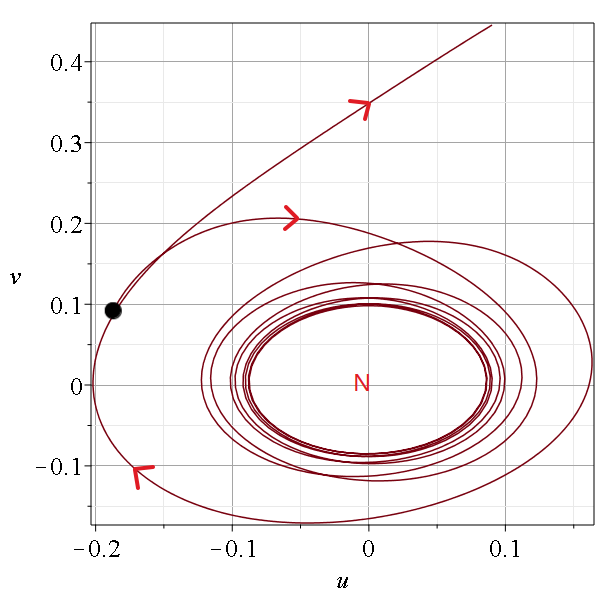}[b]\\
    \caption{A zoom in the Bohmian vortex a) in the inertial frame of reference and b) in the frame of reference of the moving nodal point  $(k=1)$ which is the center of the vortex.}
    \label{fig:vortex}
\end{figure}

The form of the spiral sections of the trajectories are shown in Fig.~\ref{fig:vortex}a in the inertial frame of reference $(x,y)$ and Fig.~\ref{fig:vortex}b in the frame of reference of the moving nodal point $k=1$. We see that the  moving particle forms loops around the nodal point $k=1$ and later it goes away. The trajectories during the approaches to $N,X$ and $Y$ without a vortex are similar to those of Figs.~\ref{fig:triplet}a,b,c.



We conclude that every "event" in the motion of a particle consists of a number of approaches to a number of nearby nodal points and the corresponding $X$ and $Y$ points during a relatively short interval of time. During that time, the trajectory may form a loop or a spiral, but it may only show some irregularity (Figs.~\ref{fig:troxia0001}a and \ref{fig:troxia03}a). The sequence of the $k's$ of nearby approaches is not the same during different events. If the approaches are close, then they produce variations in the stretching number $a$  and introduce chaos. However, distant approaches produce very small deviations in the trajectory.

\subsection{Chaos and ergodicity}

In our previous works \cite{tzemos2020ergodicity,tzemos2021role}, we studied in detail the long-time behavior of the chaotic trajectories. We found that, for a given non-zero entanglement, all chaotic trajectories have practically the same long-time point distribution. Specifically, if we consider a sufficiently large grid of square cells covering the support of the wavefunction and count how many times each trajectory passes through them, we observe a common pattern across all chaotic trajectories. 
Therefore, the  Bohmian chaotic trajectories are ergodic, since a single trajectory is sufficient to characterize the long-time chaotic behavior of the system.

\begin{figure}[H]
    \centering
    \includegraphics[width=0.4\textwidth]{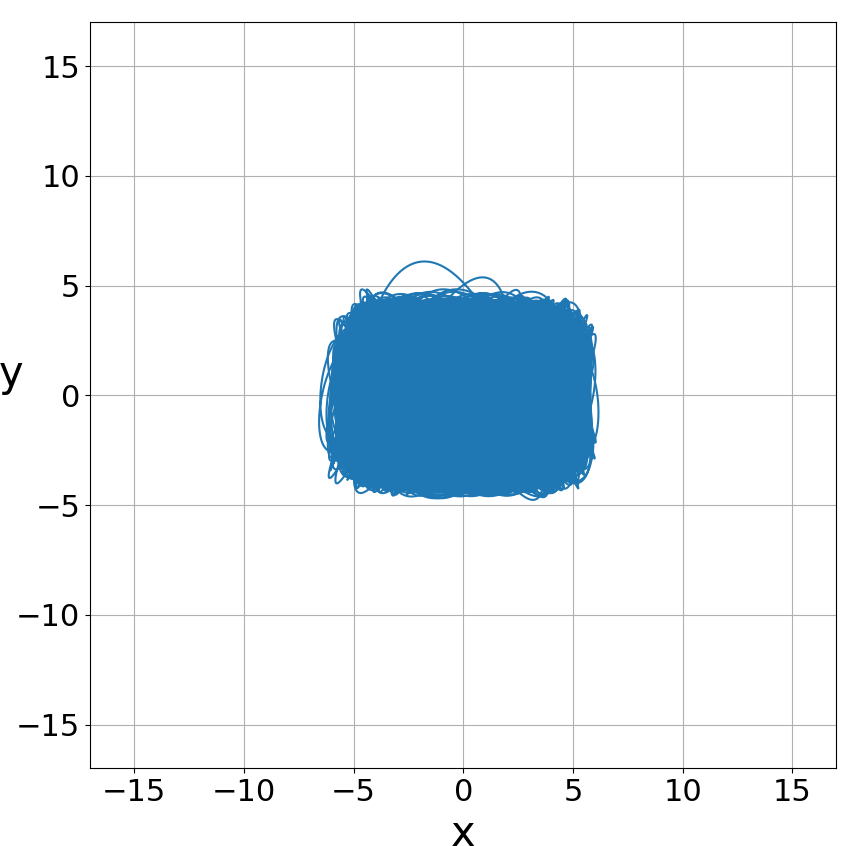}[a]
    \includegraphics[width=0.5\textwidth]{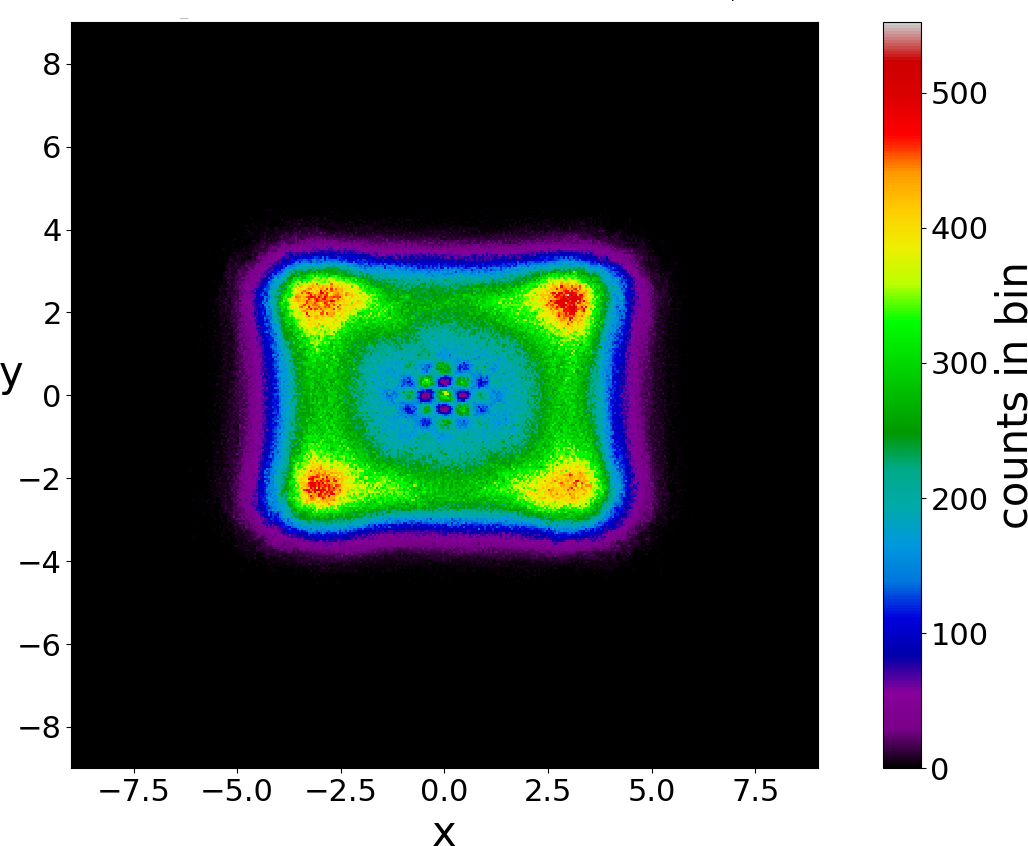}[b]\\
    \includegraphics[width=0.4\textwidth]{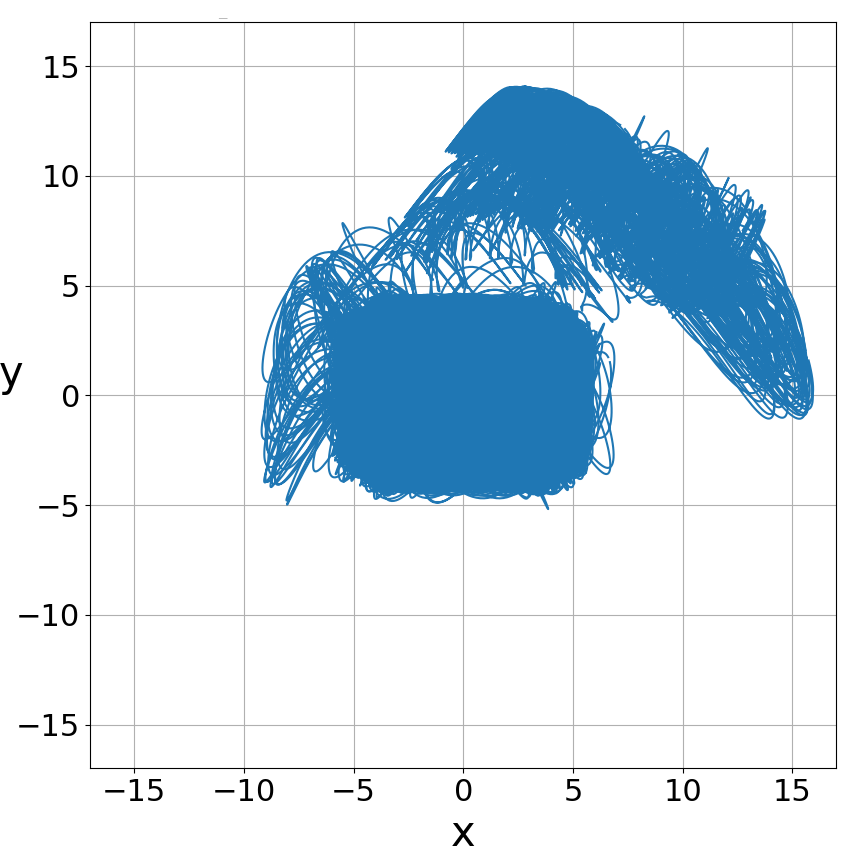}[c]
    \includegraphics[width=0.5\textwidth]{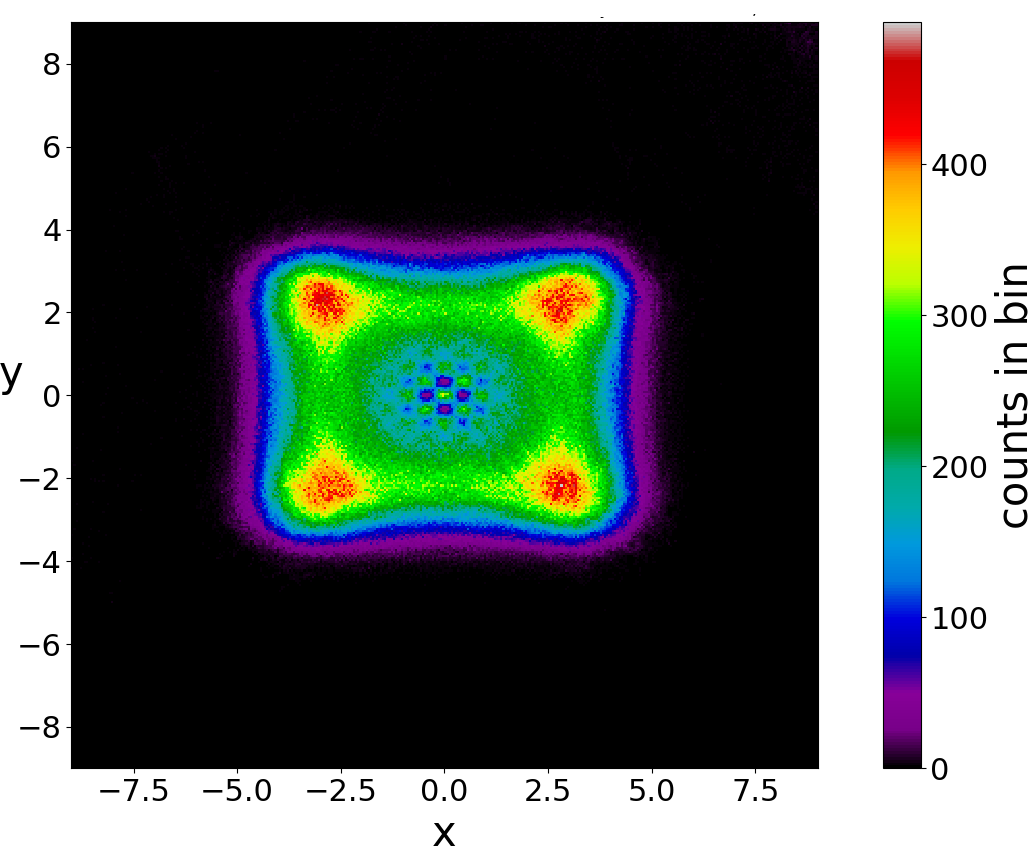}[d]
    \caption{Two different chaotic trajectories in the maximally entangled state for $\omega_x=1, \omega_y=\sqrt{3}$ (left panel): In the first case we start inside the effective support of the wavefunction ($x(0)=y(0)=2$), while in the second case we start outside ($x(0)=8, y(0)=10$), in a region where the probability of finding a particle is negligible. Once the trajectory enters the central region it remains there. The probability for the inverse process is extremely small. Their long limit colorplots (right panel) are practically the same at $t=3\times 10^5$, i.e. they are ergodic.}
    \label{fig:ergodicity}
\end{figure}

We note, however, that in classical dynamical systems, ergodicity is defined within a bounded phase space. In the quantum case, the phase space is in principle unbounded, as the support of the wavefunction extends to infinity. Nonetheless, for all practical purposes, the support is effectively bounded within a finite region around the origin. For example, working in the square region $[-6, 6] \times [-6, 6]$, we find that $|\Psi|^2 \geq 10^{-21}$, ensuring that all significant dynamics of the system  is taken into account. Outside this region, the probability of finding a particle is effectively zero. Conversely, if we consider a chaotic trajectory that starts far from the origin and integrate it for a sufficiently long time, we observe that it eventually enters the central region and remains there practically indefinitely. In Figs.~\ref{fig:ergodicity}a,c, we present two trajectories corresponding to two different initial conditions, one inside the central region and one outside, that lead to chaotic trajectories. Then we calculate their long time $(t=3\times 10^5)$ colorplots (Figs.~\ref{fig:ergodicity}b,d). The colorplots represent the number of points of a trajectory at successive steps $\Delta t=0.05$ in a dense grid of squares with side length equal to $0.05$. The darker the color of a bin in our grid, the smaller the number of points of the trajectory at this bin. {We see that although they correspond to different initial conditions they have acquired practically the same form for this given time and with a very small difference in the counts inside the bins of the grid, i.e. they are ergodic.}

\section{Ordered Trajectories}

\begin{figure}[H]
    \centering
    \includegraphics[width=0.45\textwidth]{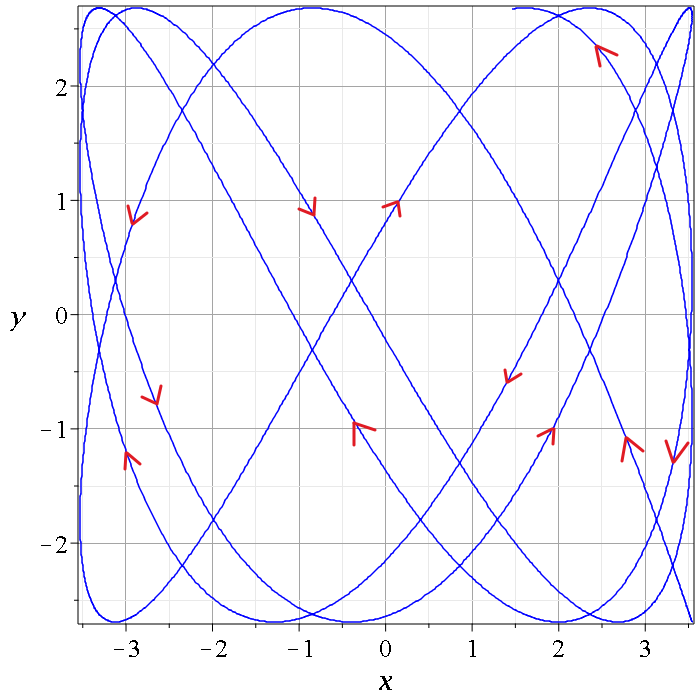}[a]\\
    \includegraphics[width=0.9\textwidth]{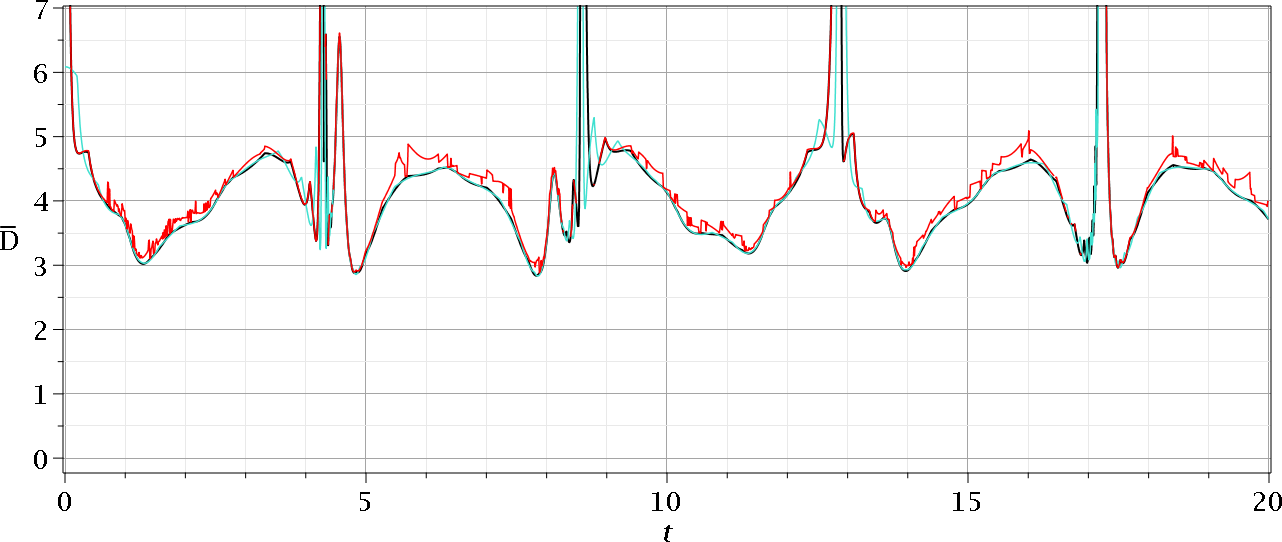}[b]\\
     \includegraphics[width=0.95\textwidth]{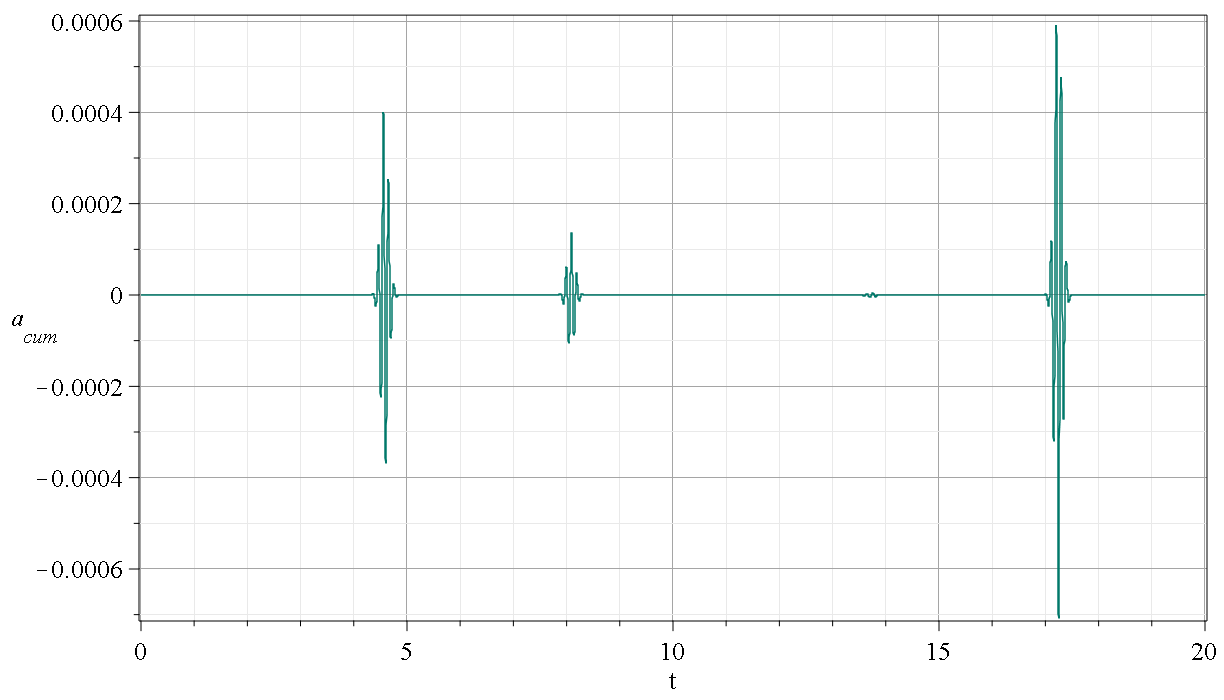}[c]
    \caption{a) An ordered trajectory in the case with  $c_2=0.001, \omega_x=1,\omega_y=\sqrt{3}$, with initial conditions $x(0)=3.54, y(0)=-2.69$ for a time interval $t=[0,20]$. b) The corresponding distances from the nearest point $N$ (black), $X$ (red) and $Y$ (green). c) The corresponding changes of the cumulative stretching number $a_{cum}=\sum a$, which are very small, of order $5\times 10^{-4}$ or smaller.}
    \label{fig:ordered}
\end{figure}

Ordered trajectories appear when the entanglement is less than its maximum value ($c_2=\sqrt{2}/2$). These trajectories are slightly deformed Lissajous figures (Fig.~\ref{fig:ordered}a). In such a case the approaches to the critical points $N,X,Y$ are very distant (we see in Fig.~\ref{fig:ordered}b that $\bar{D}>3$) and their effect is very small. The values of $a$ at these approaches are very small and of order $0.0005$ or smaller. In Fig.~\ref{fig:ordered}c we give the $a_{cum}$ at every event. We see that after every event the $a_{cum}$ goes again close to zero. Therefore the values of $\chi=\frac{1}{t}\sum a$ decrease further as time increases and tend to $LCN=0$ as $t$ tends to infinity. In fact, $\chi$ decreases approximately as $\sim t^{-1}$ (plotted on a double logarithmic scale).

A different class of ordered trajectories is found if the ratio $\omega_y/\omega_x$ is rational. If $\omega_x=s_1\omega$ and $\omega_y=s_2\omega$ the trajectory has a period $T=2\pi/\omega$. During the time $T/2$ the trajectory undergoes a number of events (approaches to the $N,X,Y$ points) and the stretching number $a$ has some variations during every event. However, at $t=T/2$ the trajectory stops (i.e. $dx/dt=dy/dt=0$) and then it retraces its  way along the same path undergoing the same events with the opposite values of $a$ until it reaches  its origin. Therefore the total sum of the values of $a$ is zero, and consequently $LCN=0$. In Appendix B we give a proof of the fact that the value of $\chi\to 0$ when $\omega_x/\omega_y$ is a rational number, while it does not go to zero when this ratio is an irrational number.

\section{Entanglement vs chaos}

In \cite{tzemos2021role}, we conducted an extensive investigation of chaos and order in  entangled qubits. More specifically, we focused on the influence of chaotic and regular dynamics on the establishment of the Born rule when the initial distribution of Bohmian particles deviates from it  \cite{valentini1991signalI,valentini1991signalII,valentini2005dynamical,durr2019typicality,lustosa2023evolution}. To this end, we  distinguished between regular and chaotic trajectories within the Born distribution for each given value of entanglement.

Our results showed that as the degree of entanglement increases, the proportion of chaotic (and ergodic) trajectories also increases, reaching  $100\%$ in the case of maximal entanglement. Consequently, in this regime, any arbitrary initial distribution of particles will evolve toward the Born distribution after a long time. However, in the case of a partially entangled state, there always exists a fraction of regular trajectories. These must be taken into account, both in terms of their proportion and of their specific locations on the Born distribution, if one aims to recover the Born rule from an arbitrary initial particle distribution.

In the present work, we take a step further by investigating how entanglement affects the degree of chaos exhibited by the chaotic trajectories themselves. Specifically, we compute the $LCN$ of a large number of chaotic trajectories across a range of entanglement values, aiming to determine the impact of entanglement on the degree of chaos. This task is well known for its numerical difficulty, due to the accumulation of round-off errors during long-time integrations, as well as the considerable computational time required to obtain reliable  estimates of the $LCN$. \footnote{In the present paper all relevant numerical computations were performed using an implementation of the (explicit) Dormand-Prince method. The absolute tolerance used was at most $a_\text{tol} = 10^{-9}$, although the minimum integration step size was always kept higher than $h = 10^{-9}$ because some orbits could prove to be extremely stiff, and would be impossible to integrate without providing a lower bound for the step size. The stretching number was calculated by re-normalizing the variational equations every $\Delta t = 0.05$.}

\begin{figure}[H]
    \centering
\includegraphics[width=0.4\linewidth]{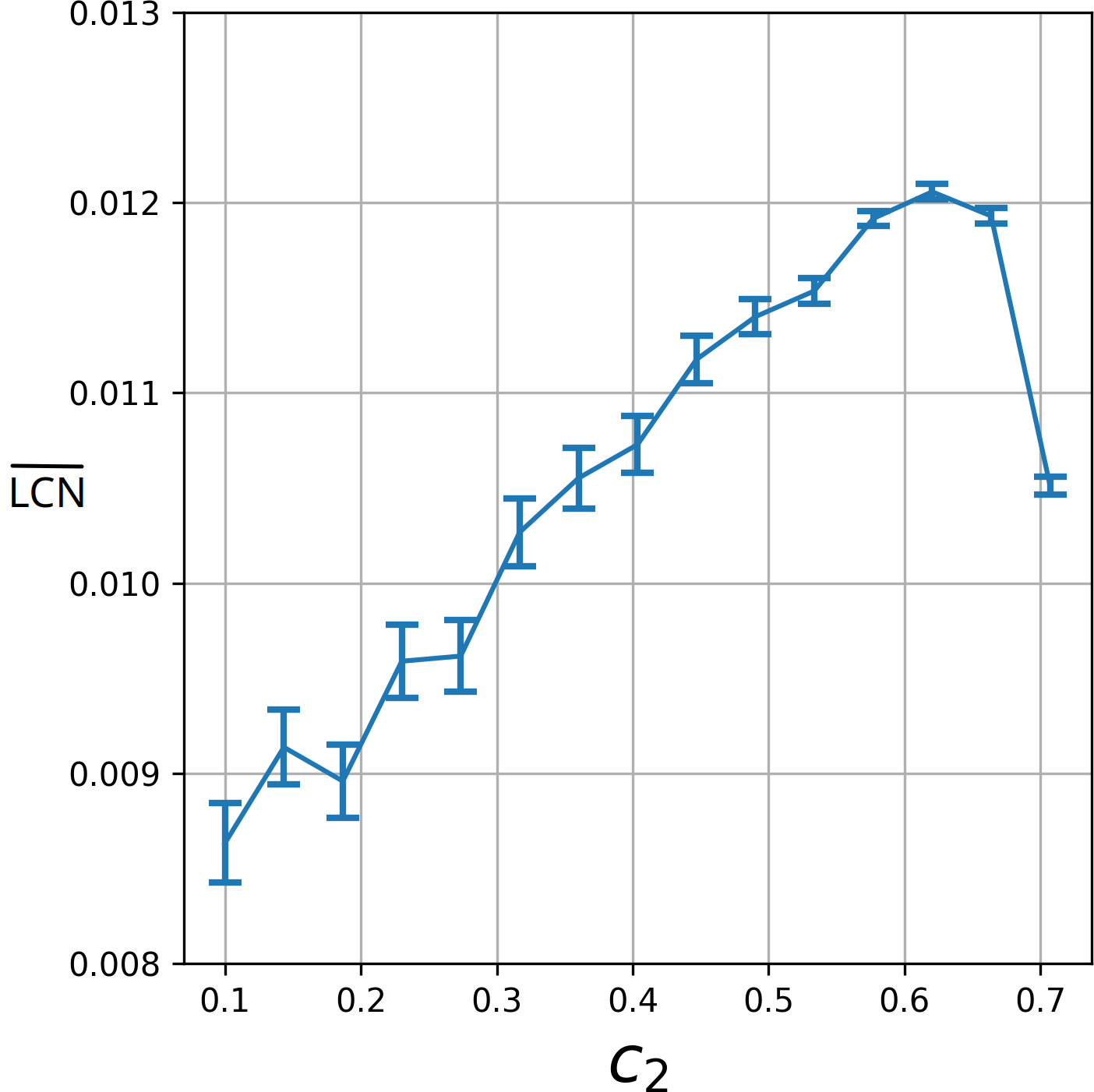}[a]
\includegraphics[width=0.4\linewidth]{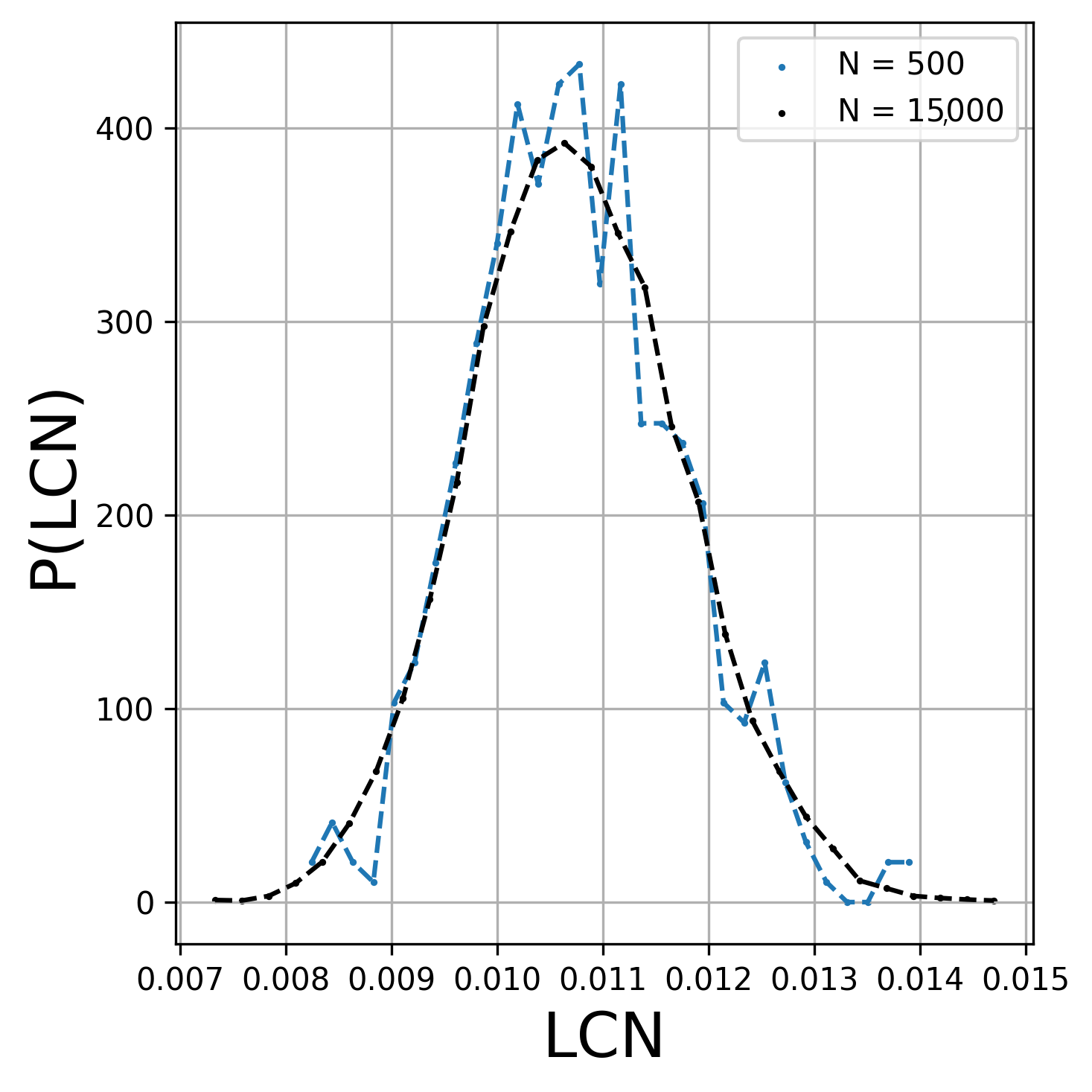}[b]
\includegraphics[width=0.4\linewidth]{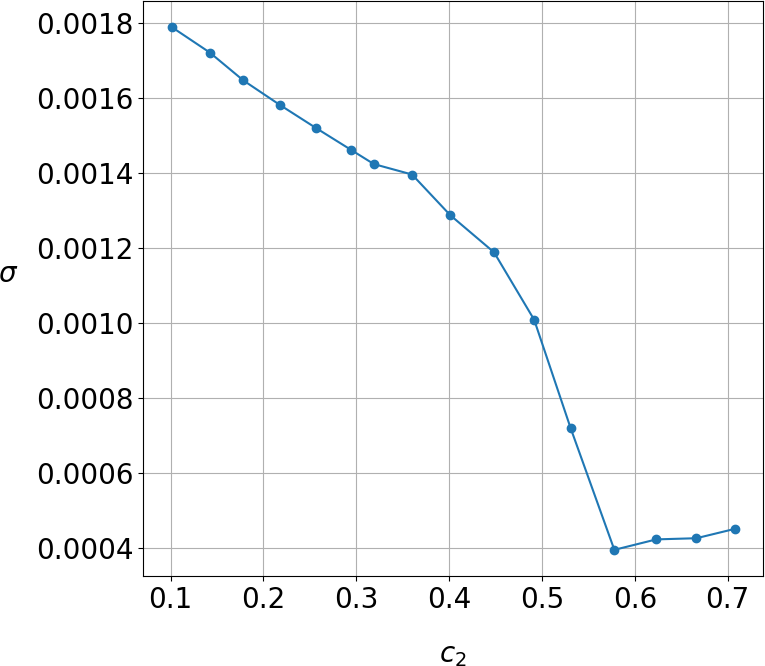}[c]
    \caption{a) The mean $LCN$ for 400 trajectories up to $t=100000$ for  $\omega_x=1, \omega_y=\sqrt{3}$ and the corresponding error bars for various values of the entanglement. b) The probability density $P(LCN)$ in the case of maximum entanglement {for $500$ (blue), and $15000$} {(black) }trajectories uniformly sampled in the grid $[-4, 4]\times[-4, 4]$ for $t=100000$. c) The standard deviations of the $LCN$ distributions in samples of 400 trajectories  as a function of the entanglement for $t=100000$.}
    \label{fig:liap1}
\end{figure}

It should be noted that in a previous study \cite{cesa2016chaotic}, a related analysis suggested no clear relationship between the degree of entanglement and chaos. We verified this in our present study. Indeed, as we show in Fig.~\ref{fig:liap1}a, the mean $LCN$ of chaotic trajectories in a set of  400 uniformly sampled trajectories in a square region $x,y=-4..4$ centered at the origin ($x=0,y=0$) does not exhibit monotonous dependence on quantum entanglement for all the  values of the latter. In fact, between $c_2\simeq 0.63$ and $c_2=0.66$ there is a maximum after which the mean $LCN$ decreases as the entanglement tends to its maximum value ($c_2=\sqrt{2}/2$). 



Furthermore, as we have shown in  \cite{tzemos2019bohmian} (see also Appendix B), in the case of commensurable frequencies all the trajectories of this system are periodic regardless the degree of entanglement. Thus, in order to observe chaos we have to work with non-commensurable frequencies. But, between two non-commensurable ratios  $\omega_x/\omega_y$ there are infinitely many commensurable ratios which lead to $LCN=0$ for every trajectory in the phase space and regardless the degree of entanglement, i.e. the value of $LCN$ has infinitely many increases and decreases as $\omega_x/\omega_y$ changes. Consequently, there is no simple relation between the entanglement and the mean value of the $LCN$.

However, we found some interesting results concerning the distribution of the values of the $LCN$.

A notable result of our calculations is that the probability density given by $P(LCN)=p(LCN)/d(LCN)$ (where $p(LCN)$ is the probability of finding a given value of $LCN$ in a small interval between $LCN$ and $LCN+d(LCN)$ (therefore $\int P(LCN)d(LCN)=1$))  for a given entanglement of a large number of trajectories  seems to have a  Gaussian form. This we  can see in Fig.~\ref{fig:liap1}b, which refers to 500 (blue) and 15000 (black) trajectories for $c_2=\sqrt{2}/2$ (maximum entanglement). We observe that as the number $N$ of the trajectories increases their distribution comes closer to a Gaussian. This Gaussian represents the trajectories along the error bar around the mean value of the $LCN$ of every entanglement.  {This behavior can be understood in terms of the Central Limit Theorem (CLT) \cite{johnson2011probability}. In chaotic systems, the $LCN$ is computed as a time average along each trajectory. Due to the sensitivity to initial conditions and the mixing properties of chaos, each trajectory effectively samples many weakly correlated regions of phase space. As a result, the finite-time LCN behaves like the average of many weakly dependent random variables, which tends toward a normal distribution according to the CLT. }


\begin{figure}[H]
    \centering
    \includegraphics[width=0.45\linewidth]{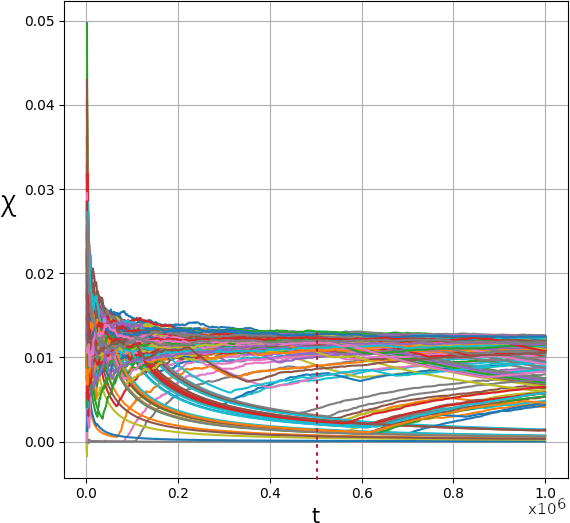}[a]
     \includegraphics[width=0.45\linewidth]{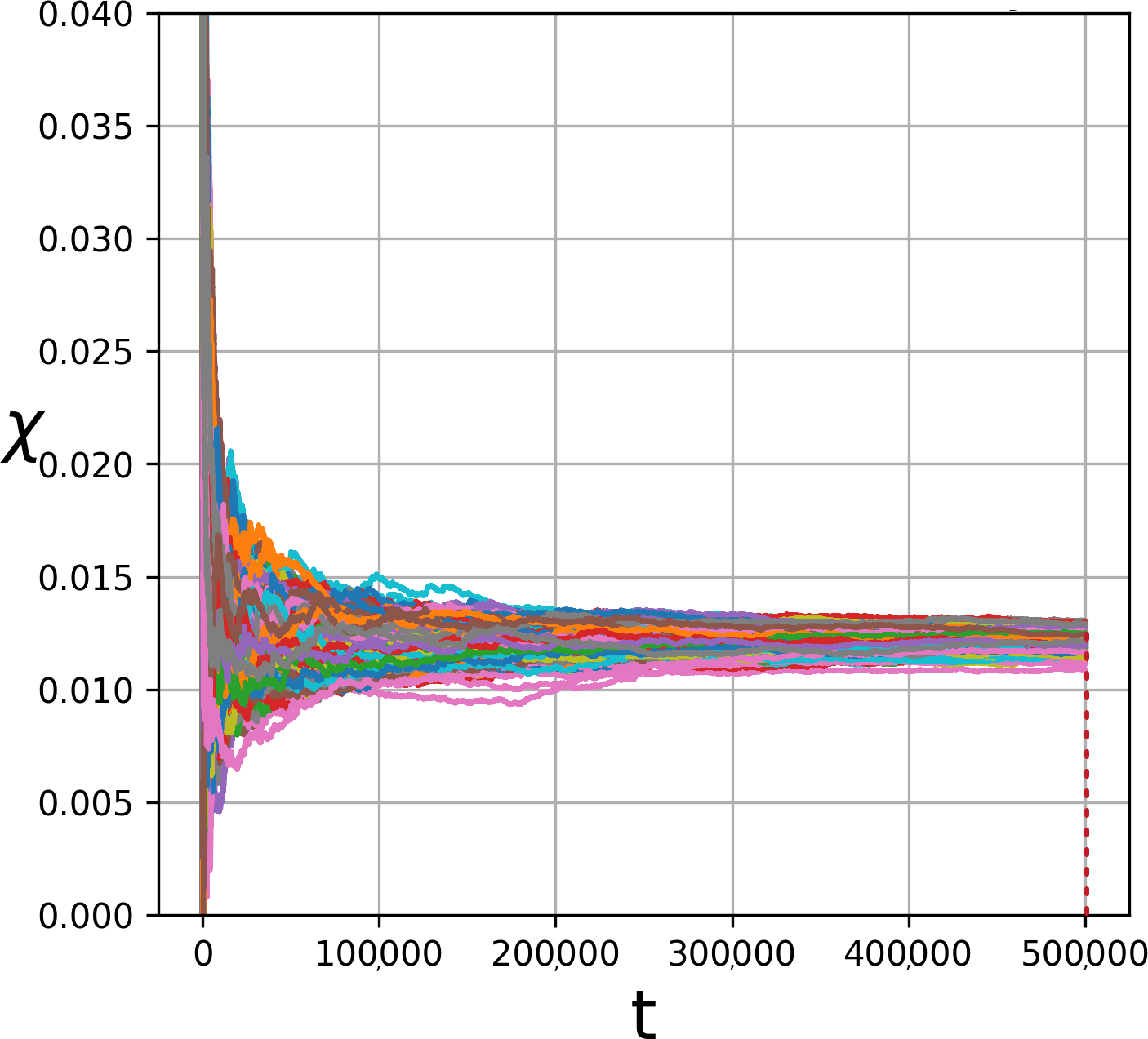}[b]
    \includegraphics[width=0.45\linewidth]{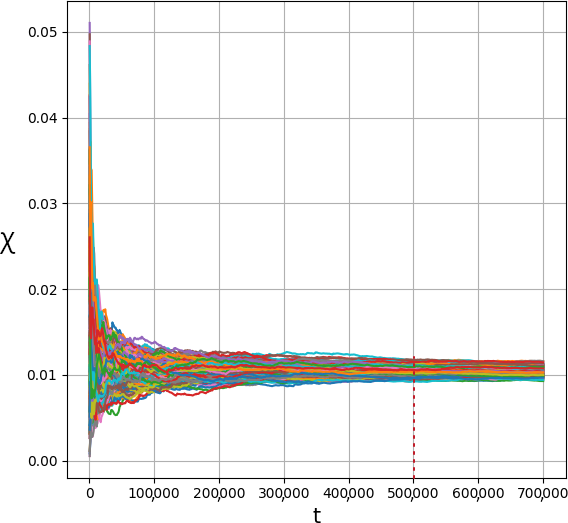}[c]
     \includegraphics[width=0.45\linewidth]{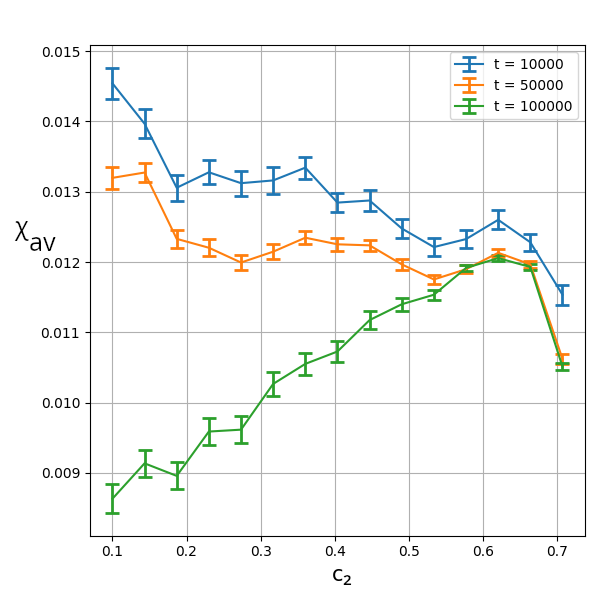}[d]
    \caption{The finite time $LCN$ of 400 trajectories uniformly sampled in the grid $[-4,4]\times[-4,4]$ for a) $c_2=0.5$, b) $c_2=0.6$ and c) $c_2=\sqrt{2}/2$. d) Average $\chi$ vs entanglement for three different integration times, $t=10^4$ (blue), $t=5\times 10^4$ (orange) and $t=10^5$ (green). In all cases $\omega_x=1,\omega_y=\sqrt{3}$. The thick lines indicate the time $t=5\times 10^5$.}
    \label{fig:liap}
\end{figure}

In Fig.~\ref{fig:liap1}c we show the standard deviation $\sigma$ as a function of $c_2$. We see that $\sigma$ decreases as the entanglement increases  to $c_2=0.59$ and for larger $c_2$ it is almost constant.  We verified that this happens in many independent repetitions of the experiment with 400 trajectories. This suggests that there is likely a relationship between entanglement and the $LCN$ convergence time. In fact, in Figs.~\ref{fig:liap}a ,b,c we have the values of the finite time $LCN$  of 400 trajectories for entanglements of $0.5,$  $0.6$ and $0.707$. We clearly see that even after one million time units in the case of Fig.~\ref{fig:liap}a, the finite time $LCN$, $\chi$,  has not reached its final value. On the other hand,  in Figs.~\ref{fig:liap}b,c where the entanglement is large and then  maximized, the various trajectories have dispersions $\Delta\chi$ of their finite time $LCN$ which decrease in time.  Based on these observations, we calculated Fig.~\ref{fig:liap}d, where we show the mean values and the error bars for the $LCN$ at different entanglement levels, and for three different observation times: $10,000$, $50,000$, and $100,000$ time units. There, we can clearly see that for small entanglements, we have large standard deviations around the mean value as well as  large differences in the mean values across the various observation times. However, as the entanglement increases, both the distance between the mean values and the size of the error bars decrease.

\begin{figure}[H]
    \centering
    \includegraphics[width=0.6\linewidth]{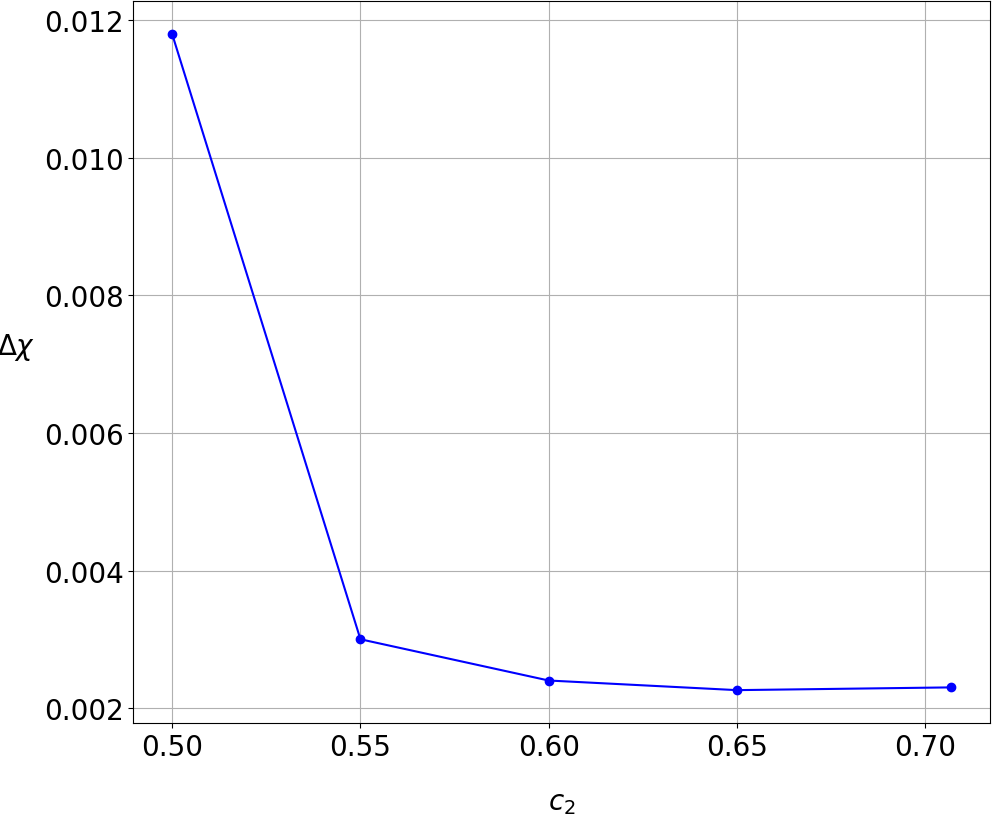}
    \caption{The range $\Delta\chi$ of the average $\chi$ at $t=5\times 10^5$ for various degrees of strong entanglement. The larger the entanglement the smaller the $\Delta\chi$.}
    \label{fig:liap3}
\end{figure}

As a consequence, the time needed for the convergence of $\chi$ to a value close to the final $LCN$ decreases with the increase of the entanglement. This can be seen in Fig.~\ref{fig:liap3}, where we give the range $\Delta\chi$ of the distribution of the values of $\chi$ at a fixed time $t=5\times 10^5$ as a function of $c_2$. We see that $\Delta\chi$ is large when $c_2=0.5,$ but decreases for larger entanglements. Therefore, for relatively large  entanglements the convergence of the values of $\chi$ is faster. This result is significant and fully consistent with our previous findings for the evolution of the probability density of this system \cite{tzemos2020ergodicity,tzemos2021role}.

\section{Conclusions}\label{sct:5}
In the present paper we studied the combined effect of the two kinds of unstable points in the Bohmian flow, the X-points (in the frame of reference of the moving node) and the Y-points (in the inertial frame of reference), in a system of two entangled qubits. The nodal points are infinitely many. They are located along a straight line and move in time going to infinity when $\omega_{xy}t=K\pi$ ($K \in \mathbb{Z}$) and coming back  from  the opposite side of infinity. Our new findings in the present paper are:


\begin{enumerate}
    \item  Every nodal point is followed by two X-points (not only one),  and between two neighboring nodal points there is a Y-point.
    \item The position of the Y-points has been given analytically, while the X-points were found only numerically.
    \item Chaos is introduced at successive points, when a particle in a relatively small interval of time approaches successively a number of nodal points and their X-points, and also the Y-points between them. A set of such approaches occurs between every two approaches to infinity. Such a set of approaches to the $N, X, Y$ is called an ``event''. We studied some characteristic events in detail.
    \item In most cases, when the particle approaches a nodal point $N$, it also approaches its X-point, which is then quite close to $N$. After that time, the particle goes further away from these points ($N$ and $X$) and approaches a Y-point between this and a neighboring nodal point. After that, the particle starts to approach the neighboring nodal and X-point. This process is repeated with approaches to a number of $N$  and $X$ points and the $Y$ points between them. The event of approaches terminates when the particle goes far from all the $N$ points.
    \item In some cases, when the particle is trapped in a very close region to the nodal point it follows a spiral motion (Bohmian vortex) around it. In this case there is a significant production of chaos.
    \item  We provided some examples of the trajectories that lead to approaches to the points $N, X, Y$, both in the inertial frame ($x$, $y$) and in the frame ($u$, $v$) around a nodal point. The points $X$ and $Y$ of the same event produce one set of variations of the stretching number and both contribute significantly in the production of chaos. {But in general the X-points are located at higher levels of the quantum potential surface than the Y-points and  the total force acting on the particle at $X$ is larger than that at $Y$.}
    \item We gave an example of an ordered trajectory which remains far from the critical points $N,X,Y$. In this case the stretching numbers are very small and the $LCN$ is zero when $\omega_x/\omega_y$ is rational. The moving particle may approach the points $N,X,Y$ (not so close, however, as in the case of non-commensurable frequencies) but its trajectory is periodic.
    \item We found that all the chaotic trajectories are ergodic, even when they start far from the central region in the $x-y$ plane. Thus, while the general form of the colorplot in a chaotic trajectory is formed in a relatively short time (we observe that it covers the full available space of the effective support of the wavefunction), the value of the $LCN$ itself needs a significantly large time to saturate on a positive number. 
    \item We found numerically that entanglement does not affect the value of the $LCN$ in a simple and unique way. However, our simulations provided evidence that the increase of entanglement decreases the convergence time of the chaotic trajectories to the final positive value of the $LCN$. 

\end{enumerate}

{The above results shed new light on the mechanisms responsible for chaos in Bohmian trajectories, highlighting the role of both X- and Y-type unstable points in the dynamics of entangled qubit systems. By analyzing how chaos arises through repeated approaches to these critical points, this study enhances our understanding of the structure and behavior of Bohmian flows. Moreover, it offers useful insight into the interplay between entanglement and chaos within the Bohmian framework, supporting our previous research on the influence of entanglement on the generation and structure of Bohmian chaos.}

{In our future research plans, we aim to explore the relationship between Bohmian chaos and the broader phenomena of quantum chaos, including signatures such as entanglement entropy growth \cite{vidmar2017entanglement}, statistical properties of energy level spacings, and spectral correlations. Understanding this relationship is basic for developing a deeper and more unified view of quantum dynamical behavior \cite{wang2004entanglement,benenti2009entanglement}.}

\section{Appendix A - Detection of the X-points}

\begin{figure}[H]
    \centering
    \includegraphics[width=0.5\linewidth]{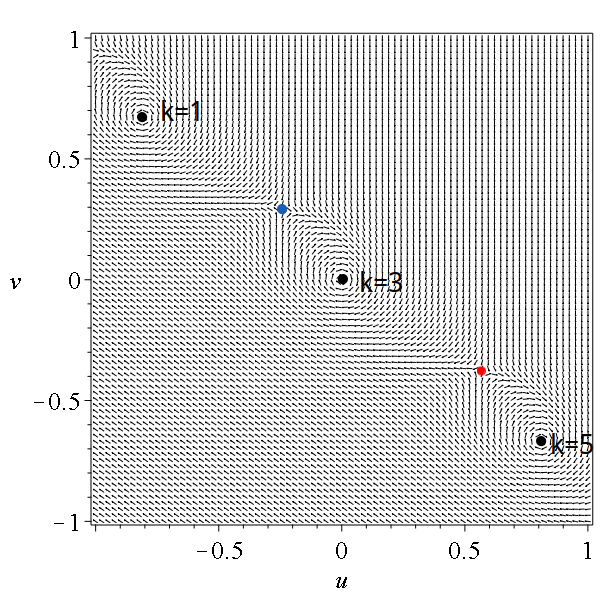}
    \caption{The Bohmian velocity field in the frame of reference of the nodal point $k=3$ for $c_2=\sqrt{2}/2$ at $t=5.8$.  All the nodal points  are shown with black dots. The X-point on the right of the node $k=3$ is marked red while that on the left is marked blue.} 
    \label{fig:append}
\end{figure}

The position of the X-points depends both on time and on the specific nodal point considered. To find their location, we first fix a time $t$ and then transform to the reference frame of a chosen nodal point $N$. In this frame, we typically find that each nodal point is accompanied by two X-points in its $(u,v)$ plane, located on each side of $N$ (Fig.~\ref{fig:append}). When we shift to the frame of the next or previous nodal point, we again observe an X-point on each side. Therefore, between any two successive nodal points, there are generally two X-points: one ahead of the first nodal point and one behind the second. When the velocities of the nodal points are small, the associated X-points are very close to each other. As the nodal velocities increase, the X-points tend to move closer to the nodal points themselves.

To understand the scattering events, we integrate the trajectory up to a fixed time using a constant time step. For the same time and step size, we store the trajectories of a large number of nodal points (i.e., for many values of $k$). At each time step, we identify the nodal point closest to the particle trajectory. We then switch to its $(u,v)$ plane, compute the corresponding X-points, determine which of them is nearest to the trajectory and store its distance from the trajectory.

The implementation of the above algorithm becomes very difficult from time to time when the nodal points acquire very large velocities before escaping to infinity. There we have NPXPCs with extremely small distance between $N$ and $X$, and the numerical convergence to the position of the X-point is slow if not impossible if we don't provide the computer with a highly accurate guess solution.

\section{Appendix B - The case of ordered trajectories}

The Bohmian equations of motion in the case of two qubits are

\begin{align}
\frac{dx}{dt} = -\frac{\sqrt{2\omega_x}}{G}\alpha_0\left[ A\cos\left(\omega_x t\right) + B\sin\left( \omega_x t\right) \right],\\
\frac{dy}{dt} = \frac{\sqrt{2\omega_y}}{G}\alpha_0\left[ A\cos\left(\omega_y t\right) + B\sin\left( \omega_y t\right) \right],
\label{eq_ddt}
\end{align}
where
\begin{align}
&\nonumber A= 2c_{1}c_{2}{{\rm e}^{2
f_{x}+2f_{y}}}\sin \left( 2(g_{x}-g_{y})\right)\\ \nonumber&
B=c_1^2e^{4f_x}-c_2^2e^{4f_y}\\&
G=2c_{1}c_{2}{{\rm e}^{2f_{x}+2f_{y}}
}\cos \left( 2(g_{x}-g_{y}) \right) +{{\rm e}^{4f_{y}}}{c_{2}}^{2
}+{{\rm e}^{4f_{x}}}{c_{1}}^{2}
\end{align}
with
\begin{eqnarray}
\nonumber f_{x}=\sqrt {2\omega_{x}} a_{0}\cos \left( \omega_{x}
\,t \right)x,\quad f_{y}=\sqrt {2\omega_{y}}a_{
0}\cos \left( \omega_{y}\,t \right) y,\\
g_{x}=\sqrt {2\omega_{x}}a_{0}\,\sin \left( \omega_{x}\,t
\right) x,\quad g_{y}=\sqrt {2\omega_{y}}a_{0}\,\sin \left( \omega_{y}\,t
\right) y
\end{eqnarray}
and $c_1^2 + c_2^2 = 1$. We have assumed $a_0 = 5/2$, $\omega_x = 1$, $\omega_y = \sqrt{3}$.

When $t=0$ we have $g_1=g_2=0$, therefore $A=0$ and $\dot{x}=\dot{y} = 0$. For $t' = -t$, we have the same $f_1$, $f_2$ but opposite $g_1$ and $g_2$. Then we find an opposite $A$ and opposite $\dot{x}$, $\dot{y}$. Therefore we describe the same trajectory in the opposite direction.

If $\omega_x/\omega_y$ is irrational, there is no point in the trajectory where both $\dot{x}$ and $\dot{y}$ are zero. However, if the fraction is rational, we have $\dot{x} = \dot{y} = 0$ not only for $t=0$ but also for every $kT/2=k2\pi/\omega, k=1,2,\dots$. In fact, at $t = T/2$, the moving particle stops and then retraces the same trajectory backwards up to $t=T$.

Then the values of $\alpha$ have the same absolute values but with opposite sign. This guarantees that the $LCN$ after one period is zero. Beyond the first period the values of $\alpha$ are the same as in the first period, but in calculating $\chi$, the denominator $t$ is now $t+T$, therefore the value of $\chi$ is smaller. The value of $\chi$ is zero at all the multiples of $t = T$ but the amplitude of $\chi$ between any two zero values become smaller and tends to zero as $t\rightarrow \infty$.

\vspace{6pt}

\section*{Acknowledgments}This work was funded by the Sectoral Development Program ($O\Pi\Sigma 5223471$) of the Ministry of Education, Religious Affairs and Sports, through the National Development Program (NDP) 2021-25. It was conducted as part of project 200/1026, supported by the Research Committee of the Academy of Athens.


\bibliographystyle{unsrt}   
\bibliography{bibliography}

\end{document}